\documentclass[aip, pof, amsmath, amssymb, amsfonts, preprint,unsortedaddress]{revtex4-2}

\usepackage{graphicx}
\usepackage{bm} 
\usepackage{mathptmx}
\usepackage{etoolbox}

\usepackage{enumerate}
\usepackage[dvipsnames]{xcolor}
\usepackage[ruled,vlined]{algorithm2e}

\newcommand{\Gp}{G^{\prime}(\omega)}
\newcommand{\Gpp}{G^{\prime\prime}(\omega)}

\usepackage{xr}
\begin{document}

\title{Sparse Regression for Discovery of Constitutive Models from Oscillatory Shear Measurements}

\author{Sachin Shanbhag${}^{*}$}%
\email{sshanbhag@fsu.edu [${}^{*}$corresponding author]}
\affiliation{Department of Scientific Computing, Florida State University, Tallahassee, FL 32306. USA}

\author{Gordon Erlebacher}%
\email{gerlebacher@fsu.edu}
\affiliation{Department of Scientific Computing, Florida State University, Tallahassee, FL 32306. USA}

\begin{abstract}
We propose sparse regression as an alternative to neural networks for the discovery of parsimonious constitutive models (CMs) from oscillatory shear experiments. Symmetry and frame-invariance are strictly imposed by using tensor basis functions to isolate and describe unknown nonlinear terms in the CMs. We generate synthetic experimental data using the Giesekus and Phan-Thien Tanner CMs, and consider two different scenarios. In the \textit{complete information} scenario, we assume that the shear stress, along with the first and second normal stress differences, is measured. This leads to a sparse linear regression problem that can be solved efficiently using $l_1$ regularization. In the \textit{partial information} scenario, we assume that only shear stress data is available. This leads to a more challenging sparse nonlinear regression problem, for which we propose a greedy two-stage algorithm. In both scenarios, the proposed methods fit and interpolate the training data remarkably well. Predictions of the inferred CMs extrapolate satisfactorily beyond the range of training data for oscillatory shear. They also extrapolate reasonably well to flow conditions like startup of steady and uniaxial extension that are not used in the identification of CMs. We discuss ramifications for experimental design, potential algorithmic improvements, and implications of the non-uniqueness of CMs inferred from partial information.
\end{abstract}


\keywords{constitutive models, large amplitude oscillatory shear, harmonic balance, machine learning, sparse regression}

\maketitle

\section{Introduction}
\label{sec:intro}

Industrially important soft materials such as polymer melts and solutions, colloidal suspensions, gels, emulsions, foams, powders, etc., are non-Newtonian fluids.\cite{larsoncf} The viscoelasticity of these complex fluids can be characterized by rheometry, in which the response of samples subjected to prescribed stress or deformation protocols is measured experimentally. Besides providing insights into material structure and behavior, such rheological measurements can guide the selection or construction of tensorial constitutive models (CMs), which mathematically describe the relationship between stress and deformation.\cite{morrison2001understanding, Larson1988} These CMs, along with conservation equations for mass and momentum, can then be incorporated into computational fluid dynamics (CFD) software to predict complex flows in process equipment such as mixers, pipe bends, contractions, injection chambers, etc., under different operating conditions.\cite{Pimenta2017, Weller1998, Favero2010} Thus, from an industrial standpoint, the development of reliable CMs for complex fluids can evade costly and time-consuming experimentation, and significantly accelerate the design and scale-up of new processes and products.

Traditionally, modelers use their experience and intuition to select a physically plausible CM. Once a CM is chosen, experimental data is used to calibrate its parameters. The advantage of this conventional workflow, especially for CMs based on microscopic physics, is that the parameters of the model are physically interpretable. The disadvantage is the problem of abundance or scarcity: Sometimes there are too many plausible CMs, while at other times there are none. Statistical tools for model selection are available to deal with the problem of abundance.\cite{Hastie2001} The problem of scarcity is more challenging. A potential recourse that has emerged in recent years is the use of rheological data to \textit{discover} the underlying CM.\cite{Brunton2016, Bortz2023, Chen2021, Lennon2023, Mahmoudabadbozchelou2024} 

Several data-driven or machine learning (ML) techniques have been proposed for the fusion of ML with CMs for rheological modeling, model identification, or calibration.\cite{Mahmoudabadbozchelou2021a, Mahmoudabadbozchelou2022a, Lennon2023a, Seryo2020, Saadat2022, John2024} A subset of these studies have focused on the discovery of CMs,\cite{Seryo2020, Lennon2023, Jin2023, Mahmoudabadbozchelou2024} which is the primary objective of this work. This enterprise is impeded by several obstacles:
\begin{enumerate}[(i)]
\item \textit{Data is limited}: Powerful ML methods thrive on digesting large quantities of information, and unearthing hidden patterns and correlations.\cite{Bishop2006} In contrast, rheological experiments are time-consuming, and the number of independent datasets $n_\text{ds}$ is typically modest  (even for well-characterized soft materials, $n_\text{ds} \approx 10 - 100$). This paucity of data makes it challenging to avoid overfitting powerful ML models, while still capturing the complexity of material behavior.
\item \textit{Designing and training neural networks is an art}: The vast majority of models that combine ML with nonlinear CMs use artificial neural networks (NNs) as universal function approximators.\cite{Mahmoudabadbozchelou2021a, Mahmoudabadbozchelou2022a, Lennon2023a, Saadat2022, Lennon2023, Mahmoudabadbozchelou2024} While the ubiquity and success of physics-inspired NNs clearly demonstrates their versatility and utility as black-box approximators, the design of suitable architectures and training protocols for optimizing parameters requires trial and error.\cite{Kaplarevic-Malisic2023} There are additional difficulties presented by (i) lack of convergence guarantees,\cite{Shin2020} (ii) stiffness of differential equations,\cite{Sharma2023, Wang2021} and (iii) sensitivity to network hyper-parameters.\cite{Novak2018, Liao2022, Wiemerslage2024} The typical strategy leans on advances in automatic differentiation by throwing a versatile optimizer like stochastic gradient descent or its variants at the problem. Unlike traditional optimization of convex objective functions, the difficulty of convergence and absence of theoretical guarantees makes designing and training NNs a dark art.
\item \textit{CMs that violate physical constraints cannot be used in CFD software}: Typically, ML is used to select, calibrate, or discover a CM that fits rheological data by using a meta-heuristic or generalized nonlinear CM as a scaffold (we adopt this approach as well). Ideally, flow fields in rheometers used to collect rheological data are spatially uniform. CMs inferred from such data might not extrapolate well to non-uniform flow fields in process equipment. Unless special care is taken, CMs inferred using ML are not guaranteed to obey physical constraints like symmetry and frame-invariance. For example, in the popular physics-informed NNs (PINNs) framework, violation of conservation laws and physical constraints are penalized, but not strictly enforced.\cite{Mahmoudabadbozchelou2021} Consequently, CMs are not guaranteed to obey physical constraints for all inputs. While such models may still be useful for \textit{interpolation} between training datasets, even minor violations of physical constraints makes them potentially unusable in CFD software. A notable exception is the approach of Lennon et al. \cite{Lennon2023} that preserves strict symmetry and frame-invariance. Consequently, we adopt certain elements of their approach in this work.

\item \textit{Inductive bias is desirable:} A purely objective data-driven approach for discovering CMs seems attractive at first glance, until we confront it with the paucity of data in rheological characterization, and the inherent nonuniqueness of inferring a general CM based on only a few measurements. More colloquially, when data is scarce, it can fit many stories, some of which may be unnecessarily complicated or incorrect. Strict adherence to physical constraints and judicious inductive biases can reduce the search space of potential CMs. An inductive bias that animates this work is parsimony, the idea that simple models are preferable to complex models. Simple models that do not violate physical constraints avoid over-fitting, which increases the odds that they extrapolate satisfactorily to novel flow and process conditions.
\end{enumerate}

\subsection{Motivation and Layout}

The primary goal of this work is to develop a protocol for the discovery of tensorial CMs from modest-sized rheological datasets that can then be embedded into CFD software. Thus, we strictly enforce physical constraints like symmetry and frame-invariance by relying on theoretical work on tensor basis functions (TBFs), which are introduced in section \ref{sec:cm_tbf_background}.\cite{Lennon2023} This strongly limits the type of nonlinear terms that can appear in a CM, and shrinks the set of spurious CMs that nevertheless fit experimental data.

We also embrace the idea of parsimony in the hope of favoring simple interpretable CMs. This is accomplished by relying on sparse regression. A general introduction is provided in section \ref{sec:sparse_regression_background}, which is specialized for the task of CM discovery in section \ref{sec:methods}. A radical departure from previous work on ML for CMs is the complete avoidance of NNs. In addition to bypassing the sensitive and opaque choices in designing and training NNs, we aim to develop a more transparent framework with the expectation that different researchers confronted with the same experimental data will reliably arrive at comparable CMs. Furthermore, we would like the method to be fast and easy to implement. Thus, our objective is to learn the CM in $\mathcal{O}(1\, \text{hr})$ on a desktop computer, using statistical techniques that do not require specialized hardware or ML training. Our overarching goal is to establish a pathway to democratize the discovery of CMs from experimental data. 

To make progress on this ambitious task, we sharpen our focus by considering a narrower problem in this work. The key elements of our approach are:

\begin{itemize}
\item \textit{Synthetic experimental data}: Oscillatory shear (OS) rheology offer a convenient experimental route for systematically exploring the linear and nonlinear rheology of materials by imposing a sinusoidal strain or stress. 
Instead of using data on real materials, we intentionally use synthetic data using well-established CMs for polymers. The advantage of using synthetic data in a proof-of-concept study of this kind is that, unlike experimental data, the `true' CM is known and can be compared against. This is helpful to characterize both the promise and inherent limitations of ML. 

\item \textit{Framework for learning CM}: Frame-invariance imposes strict limits on the structure of the time-derivatives in CMs (see Section~\ref{sec:cmOS_background}).\cite{Morozov2015} Here, we use an upper-convected derivative to model the evolution of stress,\cite{morrison2001understanding} and use data to learn the nonlinear terms of the CM. This provides a framework in which data can be incorporated by selectively learning only the unknown components of the CM. Furthermore, the unknown nonlinear term in the CM is expressed using TBFs to preserve physical constraints. 

\item \textit{Spectral method for solving nonlinear CM in OS}: We leverage a recently developed method called FLASH (Fast Large Amplitude Simulations using Harmonic balance) to obtain the periodic steady state (PSS) solution of arbitrary nonlinear differential CMs subjected to OS flow.\cite{Mittal2024a} FLASH is a spectral method that is both fast and accurate, usually by orders of magnitude compared to the standard method of numerical integration via time-stepping methods.\cite{Mittal2024, Mittal2023} It solves for the PSS solution in Fourier space, which is a natural ansatz for steady state OS flow. 

\item \textit{Sparse Regression}: We express the task of ML as a sparse regression problem.\cite{Hastie2001} In special situations where all stress components are measured, this simplifies to sparse \textit{linear} regression. Linearity puts the problem on a mathematically firm footing, where approximations and choices are automatically more transparent. Typically however, we end up with a more formidable sparse \textit{nonlinear} regression problem.\cite{Beck2013, Yang2016} FLASH is the linchpin of our strategy in this scenario. Inspired by ideas of basis pursuit,\cite{Mallat1993, Pati1993} we propose a simple greedy algorithm \cite{Hayashi2020} in Section~\ref{sec:methods_partial}.

\end{itemize}

Due to the large number of abbreviations and mathematical symbols used in this work, they are summarized for reference in the supplementary material Section~\ref{sec:nomenclature}.

\section{Background}

\subsection{Constitutive Models and Oscillatory Shear}
\label{sec:cmOS_background}

In constitutive modeling of viscoelastic liquids, we focus on the dependence of the extra stress tensor $\bm{\sigma}$ on applied deformation. It can be represented as a $3 \times 3$ matrix, or using Einstein notation as $\bm{\sigma} = \sigma_{ij} \bm{e}_i \bm{e}_j \equiv \sigma_{ij} \bm{e}_{ij}$, where $\bm{e}_1$, $\bm{e}_2$, and $\bm{e}_3$ are unit vectors in Cartesian coordinates, and the combination $\bm{e}_{ij} \equiv \bm{e}_i \bm{e}_j$ can be thought of as a $3 \times 3$ matrix whose only nonzero element is a one in the $i$th row and $j$th column.

Physics imposes important constraints on the form and evolution of $\bm{\sigma}$. Due to the conservation of angular momentum, $\bm{\sigma}$ is symmetric and has only six independent components.\cite{morrison2001understanding} Due to the principle of material frame indifference, CMs are frame invariant.\cite{Morozov2015} This severely limits the possible forms for the time derivatives in CMs. Consequently, CMs are usually formulated in terms of the upper-convected, lower-convected, and corotational derivatives, or combinations thereof.\cite{Morozov2015}  The upper-convected, lower-convected, and corotational derivatives of $\bm{\sigma}$ are given by,
\begin{align}
\overset{\triangledown}{\bm{\sigma}} \equiv \dfrac{\partial \bm{\sigma}}{\partial t} + \bm{v} \cdot \bm{\nabla \sigma } - \bm{\nabla v}^{T} \cdot \bm{\sigma} - \bm{\sigma} \cdot \bm{\nabla v},\\
\label{eqn:ucd}
\overset{\mathrel{\triangle}}{\bm{\sigma}} \equiv \dfrac{\partial \bm{\sigma}}{\partial t} + \bm{v} \cdot \bm{\nabla \sigma } + \bm{\sigma} \cdot \bm{\nabla v}^T + \bm{\nabla v} \cdot \bm{\sigma},\\
\overset{\circ}{\bm{\sigma}} \equiv \dfrac{\partial \bm{\sigma}}{\partial t} + \bm{v} \cdot \bm{\nabla \sigma } + \dfrac{1}{2} \left(\bm{\Omega} \cdot \bm{\sigma} - \bm{\sigma} \cdot \bm{\Omega}\right), 
\end{align}
respectively, where $\bm{v}$ is the velocity field, $\bm{\nabla v}$ is the velocity gradient tensor, and $\bm{\Omega}= \bm{\nabla v} - (\bm{\nabla v})^{T}$ is the vorticity tensor. The upper-convected derivative defines the upper-convected Maxwell (UCM) model, which is a simple, linear but conceptually useful model given by\cite{larsoncf} 
\begin{equation}
\stackrel{\triangledown}{\bm{\sigma}} + \dfrac{1}{\tau} \bm{\sigma} = G \dot{\bm{\gamma}},
\label{eqn:ucm_full}
\end{equation}
where $\dot{\bm{\gamma}} = (\bm{\nabla v})^{T} +  \bm{\nabla v}$ is the symmetric deformation gradient tensor. The UCM model has two material parameters: the relaxation time $\tau$, and the shear modulus $G$. For homogeneous flows such as those imposed in a rheometer, the stress field is uniform ($\bm{\nabla \sigma } = \bm{0}$), and the UCM model simplifies to
\begin{equation}
\dot{\bm{\sigma}} - \bm{\nabla v}^{T} \cdot \bm{\sigma} - \bm{\sigma} \cdot \bm{\nabla v} + \dfrac{1}{\tau} \bm{\sigma} = G \dot{\bm{\gamma}},
\label{eqn:ucm}
\end{equation}
where $\dot{\bm{\sigma}}$ denotes the time-derivative $d \bm{\sigma}/d t$. In standard OS experiments, a sinusoidal strain (or stress) is applied, and the residual periodic stress (or strain) profile once the transient response has decayed is recorded. This residual solution is called the limit cycle or the PSS solution. In this work, we focus on OS strain experiments in which a sinusoidal strain $\gamma(t) = \gamma_0 \sin \omega t$ of amplitude $\gamma_0$ and frequency $\omega$ is imposed with $\bm{e}_1$ and $\bm{e}_2$ being the shear and shear gradient directions, respectively. Then, the velocity gradient tensor $\bm{\nabla v} = \dot{\gamma} \bm{e}_{12}$ and the shear rate tensor $\dot{\bm{\gamma}} = \dot{\gamma} \bm{e}_{12} + \dot{\gamma} \bm{e}_{21}$, where the shear rate $\dot{\gamma} = \gamma_0 \omega \cos \omega t$.

This simplifies the UCM model (Equation~\ref{eqn:ucm}) in OS flow to a system of four ordinary differential equations
\begin{align}
\dot{\sigma}_{11} & + \dfrac{1}{\tau}\sigma_{11} - 2 \dot{\gamma} \sigma_{12} = 0 \nonumber\\
\dot{\sigma}_{22} & + \dfrac{1}{\tau} \sigma_{22} = 0 \nonumber\\
\dot{\sigma}_{33} & + \dfrac{1}{\tau} \sigma_{33} = 0. \nonumber\\
\dot{\sigma}_{12} & + \dfrac{1}{\tau} \sigma_{12} - \dot{\gamma} \sigma_{22} = G \dot{\gamma}.
\label{eqn:ucm_ode}
\end{align}
The other two components $\sigma_{23} = \sigma_{13} = 0$ due to the geometry of the imposed deformation. In experiments, the periodic shear, $\sigma_{12}$, and normal stress differences $N_1 = \sigma_{11} -\sigma_{22}$ and $N_2 = \sigma_{22} -\sigma_{33}$ can be measured. If we assume a stress-free initial condition $\bm{\sigma}(t=0) = \bm{0}$, then $\sigma_{33}(t) = 0$ also drops out, and we only have to track three independent components of $\bm{\sigma}$, namely $\sigma_{11}$,  $\sigma_{22}$, and $\sigma_{12}$. Consequently, from a computational standpoint, the second normal stress difference $N_2 = \sigma_{22}$ throughout this work. Since the UCM model is linear, Equation~\ref{eqn:ucm_ode} can be solved analytically to yield the PSS solution\cite{Ferry1980}
\begin{align}
\sigma_{11}(t) & = \gamma_0^2 \left[\Gp + \left(-\Gp + \dfrac{1}{2} G^{\prime}(2 \omega) \right) \cos 2 \omega t + \left(\Gpp - \dfrac{1}{2} G^{\prime\prime}(2 \omega) \right) \sin 2 \omega t \right] \notag \\
\sigma_{12}(t) & = \gamma_0 \left(\Gp \sin \omega t + \Gpp \cos \omega t \right) \notag \\
\sigma_{22}(t) & = 0,
\end{align}
where the storage modulus $\Gp = G \omega^2 \tau^2/(1 + \omega^2 \tau^2)$, and the loss modulus $\Gpp = G \omega \tau/(1 + \omega^2 \tau^2)$. Although the UCM model is a conceptually useful model of viscoelasticity, its ability to describe real materials is severely limited due to linearity. To overcome this limitation, we can consider a generalized nonlinear differential CM based on the upper-convected derivative or the \textit{generalized} UCM model as
\begin{equation}
\dot{\bm{\sigma}} - \bm{\nabla v}^{T} \cdot \bm{\sigma} - \bm{\sigma} \cdot \bm{\nabla v} + \dfrac{1}{\tau} \bm{\sigma} + \bm{F}(\bm{\sigma},  \dot{\bm{\gamma}}) = G  \dot{\bm{\gamma}},
\label{eqn:gen_UCM}
\end{equation}
where $\bm{F}(\bm{\sigma},  \dot{\bm{\gamma}})$ is a nonlinear but frame-invariant tensor function. Many popular physics-based CMs for polymer solutions and melts such as the Giesekus and the affine Phan-Thien Tanner (PTT) models fit this mold.

In the Giesekus model,\cite{Giesekus1982} $\bm{F}(\bm{\sigma},  \dot{\bm{\gamma}})$ takes a quadratic form,
\begin{equation}
\bm{F}_\text{Giesekus}(\bm{\sigma},  \dot{\bm{\gamma}}) = \dfrac{\alpha_G}{G\tau} \bm{\sigma} \cdot \bm{\sigma},
\label{eqn:F_giesekus}
\end{equation}
where the dimensionless parameter $\alpha_G \in [0, 1]$ controls nonlinearity. The Giesekus model was originally developed to describe the nonlinear viscoelastic behavior of polymer solutions in both shear and extension. Subsequently, it has been applied to other systems such as worm-like micelles\cite{Holz1999, Fischer1997, rehage2015experimental, bandyopadhyay2005effect, KateGurnon2012} and protein dispersions.\cite{kokini2000integral, Dhanasekharan2001}

Similarly, for affine flow, the nonlinear term in the exponential PTT model is\cite{Thien1977, Thien1978}
\begin{equation}
\bm{F}_\text{PTT}(\bm{\sigma},  \dot{\bm{\gamma}}) = \dfrac{f_\text{PTT}(\bm{\sigma}) - 1}{\tau} \bm{\sigma},
\label{eqn:F_PTT}
\end{equation}
where the extent of nonlinearity is controlled by a dimensionless parameter $\epsilon_\text{PTT}$ through an exponential function of the trace of $\bm{\sigma}$,
\begin{equation}
f_\text{PTT}(\bm{\sigma} ) = \exp\left(\dfrac{\epsilon_\text{PTT}}{G} \text{tr} (\bm{\sigma}) \right).
\label{eqn:fptt}
\end{equation}
The PTT model for polymeric fluids is motivated by the Lodge-Yamamoto network theory,\cite{Yamamoto1956, Lodge1956} in which cross-links can be created and destroyed. The version described by Equation~\ref{eqn:F_PTT} applies for affine flows which assumes that there is no slip between the network and the continuous medium. The dimensionless parameter $\epsilon_\text{PTT} \in [0, 1]$ models the rate at which cross-links are destroyed in the network.\cite{Thien1978} Typically, $\epsilon_\text{PTT}$ ranges from  $\approx 0.02$ for dilute polymer solutions, to $\approx 0.3$ for polymer melts. Values approaching $\epsilon_\text{PTT} \approx 1$ are considered unrealistic for describing real materials.\cite{Sibley2010} The exponential form of the nonlinearity qualitatively reproduces experimental data in strong flows,\cite{Larson1988} and multi-mode versions of the model have been used to describe the behavior of real materials.\cite{Hatzikiriakos1997, Shiromoto2010, Dietz2015}

The ordinary differential equations corresponding to the Giesekus and PTT models in OS are listed in supplementary material Section~\ref{sec:ode_models}. Both the Giesekus and PTT models reduce to the UCM model when $\alpha_G$ or $\epsilon_\text{PTT}$ are zero. However, unlike the UCM model, their OS response cannot be obtained analytically in general. The standard approach for obtaining PSS solutions of nonlinear CMs subjected to large-amplitude oscillatory shear (LAOS) is numerical integration using a suitable time-stepping method like Runge-Kutta.\cite{heath2018scientific} However, this approach can be computationally expensive because of the need for implicit methods to address numerical instability at large values of $\omega \tau$ and $\gamma_0$, and to mitigate long transients.

Recently, we developed a fast and accurate spectral method called FLASH to compute the PSS solution of any nonlinear differential CM subjected to OS strain.\cite{Mittal2024a} It is based on the technique of harmonic balance,\cite{Krack2019, Mittal2023, Mittal2024} which uses Fourier transforms to convert a system of nonlinear ordinary differential equations into a system of nonlinear algebraic equations in the Fourier coefficients. FLASH couples harmonic balance with a numerical scheme called alternating-frequency-time, which transforms the nonlinear terms in the CM to the frequency space and back during each iteration of the solver. The mathematical ideas behind this approach are presented in Section~\ref{sec:hb_flash} of the supplementary material.

FLASH offers a convenient interface to identify the OS response of arbitrary differential CMs. The methodology is not restricted to CMs based on the generalized UCM model (Equation~\ref{eqn:gen_UCM}), although that is the form used here. As input, FLASH takes in a fully parameterized differential CM, operating conditions ($\gamma_0$ and $\omega$), and a parameter $H$ which specifies the ansatz. FLASH exploits the even and odd symmetries of the normal and shear stresses, respectively, and resolves them up to the $(2H + 1)^\text{th}$ harmonic. Based on previous experience,\cite{Mittal2023, Mittal2024a} we set $H = 8$ throughout this work for simplicity, although this assumption can be easily relaxed if required. FLASH shifts the burden of setting up and solving the relevant set of harmonic balance equations from the modeler to the computer. The most attractive attributes of FLASH in this work are its accuracy and speed. In these benchmarks, FLASH typically outperforms numerical integration by 1--3 orders of magnitude.\cite{Mittal2023, Mittal2024, Mittal2024a}

\subsection{Nonlinear Constitutive Models using Tensor Basis Functions}
\label{sec:cm_tbf_background}

Consider symmetric, frame-invariant 3$\times$3 matrices $\bm{F}$, $\bm{A}$, and $\bm{B}$, where $\bm{F}(\bm{A}, \bm{B})$ is an arbitrary analytic function of $\bm{A}$ and $\bm{B}$. Using the Cayley-Hamilton theorem (see appendix \ref{app:cht}), $\bm{F}$ can be represented as a polynomial of degree two in $\bm{A}$ and $\bm{B}$,\cite{Rivlin1955, Spencer1959, Rivlin1955a, Dui2004}
\begin{align}
\bm{F}(\bm{A}, \bm{B}) = g_1 \bm{I} & +  g_2 \bm{A} + g_3 \bm{B} + g_4 \bm{A}^2 + g_5\bm{B}^2 + g_6(\bm{AB} + \bm{BA}) \notag \\
& + g_7(\bm{A}^2 \bm{B} + \bm{B} \bm{A}^2) + g_8(\bm{A}\bm{B}^2 + \bm{B}^2\bm{A}) + g_9 (\bm{A}^2\bm{B}^2 + \bm{B}^2\bm{A}^2),
\label{eqn:rivlin1}
\end{align}
where tensor dot products $\bm{A} \cdot \bm{A} = \bm{A}^2$ and $\bm{A} \cdot \bm{B} = \bm{A B}$ are equivalent to matrix multiplications. The $n_g = 9$ coefficients $\{g_i\}$ are \textit{polynomials} in the 10 invariants of $\bm{A}$ and $\bm{B}$ given by the traces of the following 10 matrices: $\bm{A}, \bm{B}, \bm{A}^2, \bm{B}^2, \bm{A}^3, \bm{B}^3, \bm{AB}, \bm{A}^2\bm{B}, \bm{A}\bm{B}^2$, and $\bm{A}^2 \bm{B}^2$. This set of 10 matrices contains all possible matrix products $\bm{A}^{i} \bm{B}^{j}$ with $i + j \leq 4$.

Equation \ref{eqn:rivlin1} provides a finite set of basis functions with which any symmetric, frame-invariant, smooth 3$\times$3 matrix function can be represented. Lennon et al.\cite{Lennon2023} recognized the importance of this result for inferring nonlinear CMs from experimental data. 
They used the generalized UCM model (Equation~\ref{eqn:gen_UCM}) as a scaffold and represented the unknown nonlinear term $\bm{F}(\bm{\sigma},  \dot{\bm{\gamma}})$ as a linear combination of polynomials in $\bm{\sigma}$ and $\dot{\bm{\gamma}}$ according to Equation~\ref{eqn:rivlin1}, leading to
\begin{equation}
\bm{F}(\bm{\sigma},  \dot{\bm{\gamma}}) = \sum_{i=1}^{n_g} g_i(\bm{L})\, \bm{T}_i(\bm{\sigma},  \dot{\bm{\gamma}}),
\label{eqn:tbf_nonlin}
\end{equation}
where $\mathbb{T} = \{\bm{T}_i\}$ is the set of TBFs, $\bm{L} = \{I_i(\bm{\sigma},  \dot{\bm{\gamma}})\}$ is the set of invariants, and $\bm{g} = \{g_i\}$ is the set of scalar coefficient functions (SCFs), with $1 \leq i \leq n_g$. The TBFs and invariants are listed in Table \ref{tab:tbf_and_invariant}. Note that only 9 invariants ($I_i$) are listed instead of the 10 anticipated,\cite{Rivlin1955, Rivlin1955a} because tr($\dot{\bm{\gamma}}$) = 0 for incompressible fluids, eliminating one of the invariants from the original set.

Lennon et al. proposed a framework called rheological universal differential equations (RUDE).\cite{Lennon2023} Universal differential equations represent a marriage of ML and differential equations, and embed universal functional approximators such as NNs within differential equations. By representing the nonlinear term using Equation~\ref{eqn:tbf_nonlin}, the goal of inferring a nonlinear CM boils down to learning the SCFs $g_1(\bm{L})$ through $g_9(\bm{L})$ by fitting experimental data. The RUDE framework accomplished this task by modeling $\bm{g}(\bm{L})$ using a deep NN. Several advantages of this approach were observed. It worked well with limited experimental data and different deformation protocols. It also  extrapolated to processing flow conditions surprisingly well.\cite{Lennon2023}

\begin{table}
\begin{center}
\begin{tabular}{l|l}
\hline
\textbf{TBFs} & \textbf{invariants} \\
\hline
$\bm{T}_1 = \bm{I}$ &  $I_1 = \text{tr}(\bm{\sigma})$\\
$\bm{T}_2 = \bm{\sigma}$ & $I_2 = \text{tr}(\bm{\sigma} \cdot \bm{\sigma})$\\
$\bm{T}_3 =  \dot{\bm{\gamma}}$ 
& $I_3 = \text{tr}( \dot{\bm{\gamma}} \cdot  \dot{\bm{\gamma}})$\\
$\bm{T}_4 = \bm{\sigma} \cdot \bm{\sigma}$ 
& $I_4 = \text{tr}(\bm{\sigma} \cdot \bm{\sigma} \cdot \bm{\sigma})$\\
$\bm{T}_5 =  \dot{\bm{\gamma}} \cdot  \dot{\bm{\gamma}}$ & $I_5 = \text{tr}( \dot{\bm{\gamma}} \cdot  \dot{\bm{\gamma}} \cdot  \dot{\bm{\gamma}})$\\
$\bm{T}_6 = \bm{\sigma} \cdot  \dot{\bm{\gamma}} +  \dot{\bm{\gamma}} \cdot \bm{\sigma}$ & $I_6 = \text{tr}(\bm{\sigma} \cdot  \dot{\bm{\gamma}})$\\
$\bm{T}_7 = \bm{\sigma} \cdot \bm{\sigma} \cdot  \dot{\bm{\gamma}} +  \dot{\bm{\gamma}} \cdot \bm{\sigma} \cdot \bm{\sigma}$ & $I_7 = \text{tr}(\bm{\sigma} \cdot \bm{\sigma} \cdot  \dot{\bm{\gamma}})$\\
$\bm{T}_8 = \bm{\sigma} \cdot  \dot{\bm{\gamma}} \cdot  \dot{\bm{\gamma}}+  \dot{\bm{\gamma}} \cdot  \dot{\bm{\gamma}} \cdot \bm{\sigma}$ & $I_8 = \text{tr}(\bm{\sigma} \cdot  \dot{\bm{\gamma}} \cdot  \dot{\bm{\gamma}})$\\
$\bm{T}_9 = \bm{\sigma} \cdot  \bm{\sigma} \cdot  \dot{\bm{\gamma}} \cdot  \dot{\bm{\gamma}} +  \dot{\bm{\gamma}} \cdot  \dot{\bm{\gamma}} \cdot \bm{\sigma} \cdot \bm{\sigma}$ & $I_9 = \text{tr}(\bm{\sigma} \cdot  \bm{\sigma} \cdot  \dot{\bm{\gamma}} \cdot  \dot{\bm{\gamma}})$\\
\hline
\end{tabular}
\end{center}
\caption{List of tensor basis functions (TBFs) and invariants of $\bm{\sigma}$ and $\dot{\bm{\gamma}}$.  \label{tab:tbf_and_invariant}}
\end{table}

In this work, we restrict ourselves to OS measurements. As mentioned previously, $\dot{\bm{\gamma}} = \dot{\gamma} \bm{e}_{12} + \dot{\gamma} \bm{e}_{21}$ has only two nonzero elements, and $\bm{\sigma} = \sigma_{11} \bm{e}_{11} + \sigma_{12} \bm{e}_{12} + \sigma_{12} \bm{e}_{21} + \sigma_{22} \bm{e}_{22}$ has only three unique nonzero elements. This further simplifies the expressions for the TBFs and invariants (see Appendix \ref{app:os_tbf}): in particular, there are only $n_l$ = 5 independent invariants, $\bm{l} = [l_1 = I_1, l_2 = I_2, l_3 = I_3, l_4 = I_4, l_5 = I_6] \subset \bm{L}$. Thus, we can attempt to learn the SCFs $\bm{g}(\bm{L})$ from OS data using this abridged set of invariants.

\subsection{Sparse Regression}
\label{sec:sparse_regression_background}

In this work, we infer $\bm{g}(\bm{L})$ (more precisely, $\bm{g}(\bm{l})$) using sparse regression, instead of NNs, for the following reasons: 
\begin{enumerate}[(i)]
\item a \textit{sparse} regression model for $\bm{g}(\bm{L})$ with only a few nonzero algebraic terms ($\mathcal{O}(10)$) is potentially easier to interpret than a NN with $\mathcal{O}(10^{3}) - \mathcal{O}(10^{5})$ parameters; 
\item the fact that SCFs are polynomials can be naturally embedded into the regression process;
\item under special conditions in which all the nonzero stress components are measured, the optimization problem reduces to sparse linear regression (SLR), which permits a fast, robust, and transparent solution (Section~\ref{sec:methods_complete}); 
\item under typical conditions in which only the shear stress is measured, the optimization problem leads to a non-convex sparse recovery problem, which can be approached with greedy algorithms (Section~\ref{sec:methods_partial}).
\end{enumerate}

We label the scenario in which all the nonzero stress components are measured as the \textit{complete information} or CI scenario. In experiments, this includes the shear stress $\sigma_{12}$, the first normal stress difference $N_1$, and the second normal stress difference $N_2$. This represents an idealized case, since most OS experiments do not measure $N_2$ due to experimental difficulties and only occasionally report $N_1$.  Therefore, for the more realistic scenario of \textit{partial information} (PI) we assume that only the shear stress ($\sigma_{12}$) is available. Specific methods for CM discovery in these scenarios are described in Section~\ref{sec:methods}. In the following, we provide a general introduction to SLR and sparse nonlinear regression (SNLR).

\subsubsection{Sparse Linear Regression}

Consider the linear least-squares (LS) solution of an over-determined linear system $\bm{M \alpha} = \bm{y}$, where $\bm{y} = [y_1, \cdots, y_n]^T$ is a vector of $n$ observations, and $\bm{M}$ is an $n \times p$ matrix of predictors with $n \gg p$. The goal of linear LS regression is to find the combination $\bm{\alpha} = [\alpha_1, \cdots, \alpha_p]^T$  of the columns of $\bm{M}$ that most closely resembles $\bm{y}$, i.e., $\bm{M \alpha} \approx \bm{y}$. Since the system is over-determined, an LS solution attempts to solve the minimization problem
\begin{equation}
\bm{\alpha}_\text{LS} = \min_{\bm{\alpha}} \Vert \bm{M \alpha - y} \Vert_{2}^{2},
\end{equation}
where $\Vert \bm{x} \Vert_q$ represents the $q$-norm of vector $\bm{x}$. The LS solution can be obtained by solving the normal equations, or using QR decomposition.\cite{heath2018scientific} Typically, most elements of the resulting $\bm{\alpha}_\text{LS}$ are nonzero, suggesting that all the columns of $\bm{M}$ are required to predict $\bm{y}$. The preference for parsimony can expressed by requiring $\bm{\alpha}$ to be sparse, with most entries equal to zero.

A popular sparsity-promoting convex optimization technique called LASSO (least absolute shrinkage and selection operator) \cite{Tibshirani1996} or basis pursuit \cite{Chen1998} combines regularization and variable selection.\cite{Tibshirani2011, Brunton2019, Hastie2001} It appends an $l_1$ regularization term to the LS objective function to penalize overfitting:
\begin{equation}
\bm{\alpha}_\text{LASSO} = \min_{\bm{\alpha}} \Vert \bm{M \alpha - y} \Vert_{2}^{2} + \mu \Vert \bm{\alpha} \Vert_{1}.
\label{eqn:lasso}
\end{equation}
The magnitude of the parameter $\mu$ simultaneously controls the strength of variable selection and regularization. As $\mu \rightarrow 0$, $\bm{\alpha}_\text{LASSO} \rightarrow \bm{\alpha}_\text{LS}$. As $\mu \rightarrow \infty$, the regularization term dominates the objective function, and leads to the trivial solution $\bm{\alpha}_\text{LASSO} = \bm{0}$. The optimal value of $\mu$, which lies somewhere between these extremes, seeks a trade-off between the need to describe the experimental observations well, the ability to pose a well-conditioned problem, and to produce a sparse solution. Often, $\mu$ is selected using cross-validation.\cite{Hastie2001, Brunton2019}

LASSO is a versatile regression technique that works for both over-determined ($n \gg p$) and under-determined ($p \gg n$) linear systems. In the latter case, it enables us to take a kitchen-sink approach by embedding all potential predictive information into the columns of $\bm{M}$, relying on LASSO to select the truly relevant set of predictors.\cite{Brunton2019} This attractive property of LASSO to identify sparse solutions is sometimes called \textit{feature selection}. Since LASSO combines variable selection and shrinkage, the nonzero elements of $\bm{\alpha}_\text{LASSO}$ are biased toward zero. This bias can be reduced by using LASSO to identify the nonzero components, and \textit{refitting} an unrestricted linear model or using LASSO again, only on the selected components.\cite{Hastie2001} Many efficient algorithms for finding $\bm{\alpha}_{\text{LASSO}}$ such as least angle regression,\cite{Efron2004} coordinate-wise optimization,\cite{Friedman2010} and basis pursuit\cite{Chen1998} have been proposed. The cost per iteration is on the same order as linear LS regression, i.e., $\mathcal{O}(n^3 + np^2/2)$.\cite{Efron2004}

\subsubsection{Sparse Nonlinear Regression}

Consider $y = f(z, \bm{\alpha}) + \text{noise}$, where $y \in \mathbb{R}$ is a noisy response or output variable, $z \in \mathbb{R}$ is a covariate or input variable, and $\bm{\alpha} \in \mathbb{R}^{p}$ is a vector that parameterizes the nonlinear model 
$f(z, \bm{\alpha}):\mathbb{R} \times \mathbb{R}^{p} \rightarrow \mathbb{R}$. Given a set of observations $\{y_i\}_{i=1}^{n}$, corresponding to inputs $\{z_i\}_{i=1}^{n}$, the sparse nonlinear recovery problem may be posed as minimization of an LS loss function,
\begin{equation}
\bm{\alpha}^{*} = \min_{\bm{\alpha}} \sum_{i=1}^{n} \left( y_i - f(z_i, \bm{\alpha}) \right)^2, \text{ subject to } \Vert \bm{\alpha} \Vert_{0} \le k,
\label{eqn:sparse_nonlin_gen}
\end{equation}
where $k \ll p$ is a positive integer that controls the number of nonzero elements in $\bm{\alpha}$ via the 0-norm $\Vert \bm{\alpha} \Vert_{0}$. Even if the $l_0$ regularization penalty in Equation~\ref{eqn:sparse_nonlin_gen} is replaced by an $l_1$ regularization term (similar to Equation~\ref{eqn:lasso} in LASSO regression), we have to contend with a non-convex optimization problem because $f(\cdot)$ is nonlinear.\cite{Yang2016} Compared to sparse linear regression, this is a much harder problem. Nevertheless, substantial theoretical progress has been made in establishing sufficient and necessary conditions for the recovery of sparse or structured signals from relatively few nonlinear observations.\cite{Blumensath2013, Beck2013} 

\medskip

Different algorithms inspired by sparse linear regression for compressed sensing applications such as iterative hard thresholding, and basis pursuit have been adapted for sparse nonlinear regression.\cite{Blumensath2008a, Blumensath2008}  Each iteration of a typical thresholding algorithm alternates between a regular gradient descent step followed by a hard or soft thresholding step, in which all but the $k$ largest elements of $\bm{\alpha}$ by magnitude are set or pushed toward zero.\cite{Beck2013, Blumensath2013, Yang2016, Yuan2018} Since the sparse nonlinear recovery problem is NP-hard,\cite{Natarajan1995} an alternative approach inspired by basis or matching pursuit,\cite{Mallat1993, Pati1993} adopts a greedy approach in which locally optimal choices are proposed at each iteration to obtain an approximate solution. In this class of methods,\cite{Blumensath2008, Beck2013, Das2014} promising features of $\bm{\alpha}$ are iteratively marked for inclusion depending on how effective they are in reducing the loss function locally.


In the following, we customize SLR for the CI scenario in Section~\ref{sec:methods_complete} and SNLR for the PI scenario in Section~\ref{sec:methods_partial}.

\section{Methods}
\label{sec:methods}

In this work, we first generate synthetic data with different combinations of strain amplitude and frequency ($\gamma_0$, $\omega$) using the Giesekus and PTT models. We set the linear viscoelastic parameters $G$ and $\tau$ equal to one, so that they set the units of stress and time, respectively. We set the nonlinear parameters $\alpha_G = 0.3$ for the Giesekus model and $\epsilon_\text{PTT} = 0.3$ for the PTT model. We generate $n_\text{ds} = 12$ synthetic datasets from three different frequencies $\omega \tau = \{10^{-1}, 1, 10^{1}\}$ that span the relaxation time ($\tau = 1$), and four different strain amplitudes $\gamma_0 = \{0.5, 1, 2, 5\}$, focusing on the medium and large amplitude regimes to elicit a vigorous nonlinear response. We use FLASH with $H = 8$ to compute the PSS solution up to the $2H + 1 = $ 17th harmonic. For the range of parameters and operating conditions explored in this work, the magnitude of the intensity of the 17th harmonic relative to the 1st harmonic is less than $10^{-4}$. For each choice of $(\gamma_0$, $\omega)$, the periodic solution ($N_1 = \sigma_{11} - \sigma_{22}$, $N_2 = \sigma_{22} - \sigma_{33} = \sigma_{22}$, and $\sigma_{12}$) is obtained over one cycle on a uniform temporal grid with $n_t = 2^6$ points. 

Before we present specific methods to infer the SCFs $\bm{g}(\bm{l})$ for the CI and PI scenarios, we discuss the approach to specify a set or dictionary of potential features with which to capture the mathematical dependence of the SCFs ($\bm{g}$) on the invariants ($\bm{l}$). This step is required for both linear and nonlinear sparse regressions.



\subsection{Dictionary of Polynomial Features}
\label{sec:PF_methods}


\begin{table}
\begin{center}
\begin{tabular}{cccl}
\hline
$d$ & $n_b$ & $n_{\alpha}$ & PFs\\
\hline
0 & 1 & 9 & $\bm{B}_0 = \{1\}$\\
1 & 6 & 54 & $\bm{B}_1 = \bm{B}_0 \cup \{l_i\}$ \\
2 & 21 & 189  & $\bm{B}_2 = \bm{B}_1 \cup \{l_i l_j\}$\\
3 & 56 &  504 & $\bm{B}_3 = \bm{B}_2 \cup \{l_i l_j l_k\}$ \\
4 & 126 &  1134 &  $\bm{B}_4 = \bm{B}_3 \cup \{l_i l_j l_k l_m\}$\\
\hline
\end{tabular}
\end{center}
\caption{Polynomial approximation of SCFs $\{g_i(l_1, l_2, l_3, l_4, l_5)\}_{i=1}^{9}$ with degree $d$ for $n_l = 5$. The number of unknown coefficients, $n_\alpha = 9 n_b$. The set $\{l_i l_j \cdots l_q\}$ denotes all unique products of invariants with  $1 \leq i, j, \cdots, q \leq n_l$. Thus, $\{l_i l_j\} = [l_1^2, l_1 l_2, l_1 l_3, \cdots, l_1 l_5, l_2^2, l_2 l_3, \cdots l_2 l_5, l_3^2, \cdots, l_5^2]$. \label{tab:PFs}}
\end{table}


As mentioned in section \ref{sec:cm_tbf_background}, for OS flow the SCFs $g_i(\bm{l})$ are \textit{polynomials} in $n_l = 5$ independent invariants, $\bm{l} = [l_1=I_1, l_2 = I_2, l_3 = I_3, l_4 = I_4, l_5=I_6]$. Mathematically, we approximate $g_i(l_1, l_2, l_3, l_4, l_5)$ for $i \in [1, n_g = 9]$ by restricting the representation of the polynomial features up to a specified degree. Let $d$ be the degree of polynomial approximation for $\bm{g}$, and let $\bm{B}_{d} = \{B_j\}_{j=1}^{n_b}$ denote the set of all unique polynomial combinations of the invariants with degree less than or equal to $d$. Let us make this notion more concrete by considering a toy example with $n_l = 2$ invariants, $l_1$ and $l_2$. If $d = 2$, then the set of $n_b = 6$ polynomial features (PFs) is $\bm{B}_2 = \{1, l_1, l_2, l_1^2, l_2^2, l_1 l_2\}$. The elements of $\bm{B}_{d}$ constitute $n_b$ PFs, $\{B_j(\bm{l})\}_{j=1}^{n_b}$. As $d$ and $n_l$ increase, $n_b$ also increases. In general, the number of PFs $n_b$ can be expressed using the binomial coefficient as
\begin{equation}
n_b = {}^{n_l + d}C_{d} = \dfrac{(n_l + d)!}{n_l!\, d!}.
\label{eqn:nb}
\end{equation}
In the toy example, $n_l = 2$ and $d = 2$. Thus, there are $n_b = 4!/(2!\, 2!) = 6$ PFs in $\bm{B}_2$. We approximate the $i$th SCF as 
\begin{equation}
g_i(\bm{l}) = \sum_{j=1}^{n_b} \alpha_{i,j} B_j(\bm{l}), \quad \quad \text{ where } 1 \leq i \leq n_g = 9.
\label{eqn:glamBasis}
\end{equation}
The coefficients ($\alpha_{i,j}$) are the $n_{\alpha} = n_g n_b$ unknowns. In the toy example, this implies,
\begin{equation}
g_i(l_1, l_2) = \alpha_{i,1} + \alpha_{i,2} l_1 + \alpha_{i,3} l_2 + \alpha_{i,4} l_1^2 + \alpha_{i,5} l_2^2 + \alpha_{i,5} l_1 l_2.
\end{equation}
Discovery of the CM from experimental data is thus reduced to determining the $n_\alpha$ coefficients.

For OS experiments, $n_b = $ 6, 21, or 56 and $n_{\alpha} = $ 54, 189, and 504, for $d = $ 1, 2, or 3, respectively (see Table \ref{tab:PFs}). Although this increase with $d$ may seem explosive, a relatively small $d$ often suffices to describe highly nonlinear behavior because most of the invariants and TBFs are themselves nonlinear. As we shall see shortly, $d = 0$ is sufficient to describe the Giesekus model. Even the exponential PTT model can be approximated by $d = 1$, under suitable conditions. 

Thus, the number of unknown coefficients $n_\alpha$ is modest, especially when compared to the number of parameters in deep NNs. In sparse regression, most of these coefficients turn out to be zero. In what follows, we use the symbol $\bm{\alpha}$ to denote the vector in which these coefficients are stacked. Depending on the context, these elements are indexed as $\alpha_{i, j}$ where $i \in [1, n_g]$ and $j \in [1, n_b]$, or simply as $\alpha_k$ where $k = (i-1) n_b + j \in [1, n_\alpha]$. Computationally, this is equivalent to reshaping $\bm{\alpha}$ into a matrix or a vector.

\begin{figure}
\begin{center}
\includegraphics[scale=0.7]{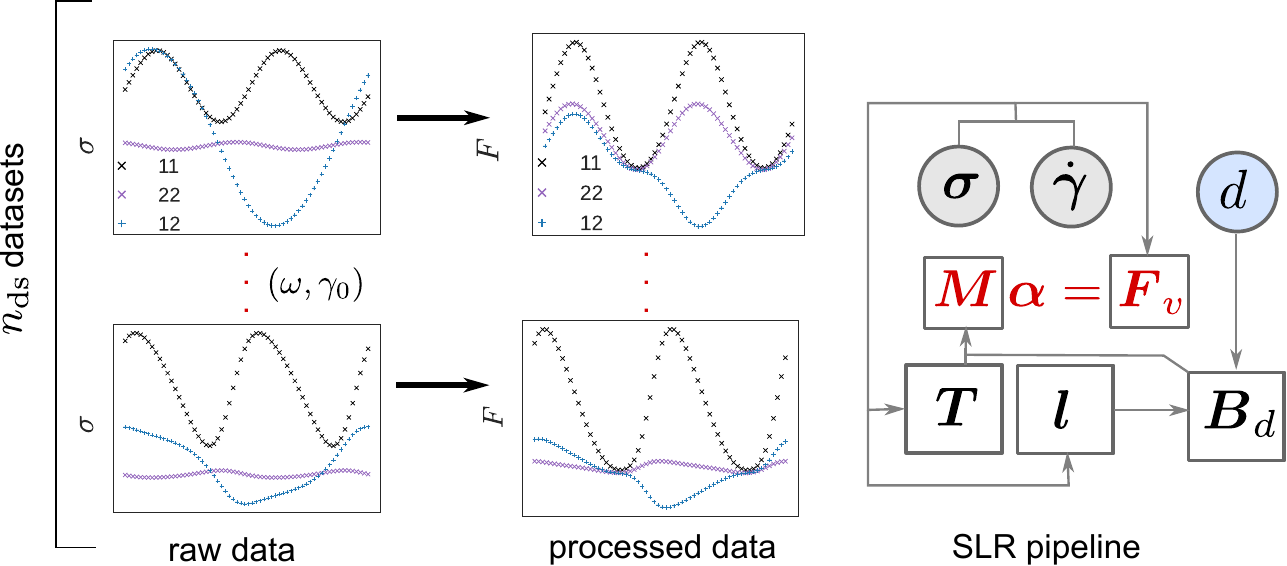}
\end{center}
\caption{\textbf{Schematic of SLR in the CI scenario}. Raw data is processed to obtain the nonlinear term $\bm{F}$ in the generalized UCM model. The TBFs and invariants are evaluated using $\bm{\sigma}$ and $\dot{\bm{\gamma}}$. Once the degree $d$ of polynomial approximation is selected, the set of PFs $\bm{B}_d$ can be computed to formulate a linear system of equations in the unknown coefficients ($\bm{\alpha}$).
 \label{fig:ci_schema}}
\end{figure}

\subsection{Inference under Complete Information}
\label{sec:methods_complete}

Let begin with the simpler case of CI in which raw data consists of the $n_{\sigma}$ = 3 nonzero stress components, $\sigma_{11}$, $\sigma_{22}$, and $\sigma_{12}$ reported at the $n_\text{ds}$ different settings of frequency and strain amplitude (see Figure~\ref{fig:ci_schema}). In an actual experiment, this implies that, in addition to shear stress, both $N_1$ and $N_2$ are also measured. For each dataset in this scenario, we report the PSS profiles obtained via FLASH at $n_t = 2^6$ equispaced points over a period of oscillation, $t \in [0, 2\pi/\omega]$.  Thus, all the components of $\bm{\sigma}$ are effectively known, since components that are not directly specified can either be inferred from symmetry ($\sigma_{21} = \sigma_{12}$) or set equal to zero (all other components).

Since $\dot{\bm{\gamma}}$ and $\bm{\sigma}$ are known, we can substitute them into the generalized UCM model (Equation~\ref{eqn:gen_UCM}) to isolate and evaluate the nonlinear term $\bm{F}(\bm{\sigma},  \dot{\bm{\gamma}})$ for each dataset $p \in [1, n_\text{ds}]$,
\begin{equation}
\bm{F}^{(p)}(\bm{\sigma}^{(p)},  \dot{\bm{\gamma}}^{(p)}) =  G  \dot{\bm{\gamma}}^{(p)} - \left(\stackrel{\triangledown}{\bm{\sigma}}^{(p)} + \dfrac{1}{\tau} \bm{\sigma}^{(p)} \right).
\label{eqn:Fnl_CI_isolate}
\end{equation}
$\bm{F}$ is symmetric and frame-invariant like $\bm{\sigma}$, and for OS flow it has $n_{\sigma}$ = 3 independent components -- $F_{11}(t)$, $F_{12}(t)$, and $F_{22}(t)$ -- that are also periodic functions. The time-derivative $\dot{\bm{\sigma}}$ required to obtain $\bm{F}^{(p)}$ in Equation~\ref{eqn:Fnl_CI_isolate} is computed by first transforming $\bm{\sigma}(t)$ into Fourier space to obtain $\hat{\bm{\sigma}}(\omega)$, and truncating beyond the $(2H+1)$th harmonic. We represent such truncated Fourier coefficients using a ``hat". Next, we take the inverse Fourier transform of $i \omega \hat{\bm{\sigma}}(\omega)$ to obtain $\dot{\bm{\sigma}}(t)$. Both these steps are implemented efficiently using fast Fourier transforms (FFT).

The TBFs $\bm{T}(t)$ and invariants $\bm{l}(t)$ are also periodic and can also be evaluated using $\dot{\bm{\gamma}}$ and $\bm{\sigma}$ using equations described in Appendix~\ref{app:os_tbf}. Thus, $\bm{F}(t)$, $\bm{T}(t)$, and $\bm{l}(t)$ are periodic functions that can be provided for each data set using $\bm{\sigma}(t)$ and $\dot{\bm{\gamma}}(t)$. The nonlinear term $\bm{F}$ can then be approximated using Equations \ref{eqn:tbf_nonlin} and \ref{eqn:glamBasis} as 
\begin{equation}
\bm{F}(t; \bm{\sigma},  \dot{\bm{\gamma}}) = \sum_{i=1}^{n_g} g_i(\bm{l}(t))\, \bm{T}_i(t) \approx \sum_{i=1}^{n_g} \sum_{j=1}^{n_b} \alpha_{i,j} B_j(\bm{l}(t)) \, \bm{T}_i(t).
\label{eqn:Falpha}
\end{equation}
When the polynomial approximation degree $d$ is assumed, the PFs
$\bm{B}_{d} = \{B_j\}_{j=1}^{n_b}$ are specified. Equation \ref{eqn:Falpha} is a linear regression problem in the unknown coefficients $\bm{\alpha}$, since all the other quantities can be evaluated using $\dot{\bm{\gamma}}$ and $\bm{\sigma}$.

Let us now mathematically define the loss function that we seek to minimize. The loss function can be expressed as the sum of squared residuals (SSR) over all the observations:
\begin{equation}
\chi^{2}(\bm{\alpha}) =  \dfrac{1}{2} \sum_{p=1}^{n_\text{ds}} \sum_{k=1}^{n_t} \bigg\Vert \bm{F}^{(p)}(t_k) - \sum_{i=1}^{n_g} \sum_{j=1}^{n_b} \alpha_{i,j} B_j^{(p)}(t_k) \, \bm{T}_i^{(p)}(t_k) \bigg\Vert^{2}.
\label{eqn:loss}
\end{equation}
It involves summation over five indices: $i \in [1, n_g = 9]$ over the SCFs, $j \in [1, n_b]$ over the PFs, $k \in [1, n_t = 2^6]$ over the discrete time points, and $p \in [1, n_\text{ds}]$ over the different datasets $(\omega^{(p)}, \gamma_{0}^{(p)} )$.  In Equation~\ref{eqn:loss}, the squared norm of a 3 $\times$ 3 symmetric matrix $\bm{A}$ with three independent nonzero elements $A_{11}$, $A_{22}$ and $A_{12}$ is defined as $\Vert \bm{A} \Vert^{2} = A_{11}^2 + A_{22}^2 + A_{12}^2$. The quadratic dependence of $\chi^{2}(\bm{\alpha})$ on $\bm{\alpha}$ leads to the linear LS problem $\bm{M \alpha} = \bm{F}_v$, where $\bm{\alpha}$ is an $n_{\alpha}$ dimensional vector of unknown coefficients, $\bm{F}_v$ is a vector with $n_{F_v} = n_\text{ds}  n_{\sigma} n_t$ elements of ``observations'' $\bm{F}^{(p)}(t_k)$, and $\bm{M}$ is a $n_{F_v} \times n_{\alpha}$ matrix whose elements are obtained from the invariants, PFs, and TBFs.

LASSO can then be used to seek a sparse solution,
\begin{equation}
\bm{\alpha}_\text{LASSO} = \min_{\bm{\alpha}} \chi^2_{\text{LASSO}}(\bm{\alpha}, \mu),
\end{equation}
where an $l_1$ regularization term is appended to the loss function,
\begin{equation}
\chi^2_{\text{LASSO}}(\bm{\alpha}, \mu) = \chi^2(\bm{\alpha}) + \mu \sum_{i=1}^{n_g} \sum_{j=1}^{n_b} \lvert \alpha_{i, j} \rvert.
\label{eqn:chilasso}
\end{equation}
We find the optimal value of $\mu$ by 5-fold cross-validation using the built-in function \texttt{LassoCV} from the \texttt{linear\_model} module of the Python machine learning library \texttt{scikit-learn} version 1.11.\cite{Pedregosa2011} This implementation uses coordinate descent to fit the unknown coefficients and a duality gap calculation to control convergence.\cite{Friedman2010, Kim2008}

LASSO automatically identifies the most important PFs and sets $\alpha_{i, j} = 0$ for other features.\cite{Tibshirani2011} Thus, even if we specify a large number of redundant PFs, LASSO isolates the handful of features that primarily explain the observations. This helps to investigate the source of nonlinearity when the underlying CM is unknown. The algorithm used for the CI scenario is summarized as Algorithm \ref{algo:ci}. As input, it takes in experimental data which includes the operating conditions $(\omega, \gamma_0)$ and the components of $\bm{\sigma}(t)$ for each dataset. The degree of polynomial approximation $d$ is also specified as input. The raw experimental data are processed to obtain the nonlinear term $\bm{F}(\bm{\sigma}, \dot{\bm{\gamma}})$ which eventually forms the right-hand side $\bm{F}_v$ of the linear system. The matrix $\bm{M}$ requires TBFs and invariants that are reconstructed from experimental data, and the set of polynomial features $\bm{B}_d$, which are constructed using the invariants and $d$. The optimal value of the regularization parameter $\mu$ is found using 5-fold cross-validation, and the sparse solution $\bm{\alpha}^{*}$ is obtained by minimizing $\chi^{2}_\text{LASSO}$ in Equation \ref{eqn:chilasso}.

\begin{algorithm}[h!]
\SetAlgoLined
\SetArgSty{textrm}
\DontPrintSemicolon
\KwIn{$(\omega^{(p)}, \gamma_{0}^{(p)} )$ and $\{ \sigma_{11}^{(p)}(t_k), \sigma_{22}^{(p)}(t_k), \sigma_{12}^{(p)}(t_k)\}_{k=1}^{n_t}$ for $p = 1, \cdots, n_\text{ds}$ datasets, degree of polynomial approximation $d$.}
\KwResult{vector $\bm{\alpha}^{*}$ to approximate SCFs $\bm{g}(\bm{l})$}
\BlankLine
\tcc{Loop over datasets}
\For{$p \gets 1$ \KwTo $n_\text{ds}$}{
    use FFT and inverse FFT to compute the time-derivatives $\{ \dot{\sigma}_{11}^{(p)}(t_k), \dot{\sigma}_{22}^{(p)}(t_k), \dot{\sigma}_{12}^{(p)}(t_k)\}_{k=1}^{n_t}$\;
    isolate and evaluate the nonlinear term $\bm{F}^{(p)}$ (Equation~\ref{eqn:Fnl_CI_isolate})\;
    furnish TBFs $\mathbb{T}^{(p)}$, invariants $\bm{l}^{(p)}$, and PFs $B_d(\bm{l}^{(p)})$\;   
}
assemble data into a linear system $\bm{M} \bm{\alpha} = \bm{F}_v$\;
use 5-fold cross-validation to determine regularization parameter $\mu$\;
solve for $\bm{\alpha}^{*}$ using LASSO regression\;
 \caption{SLR to infer $\bm{\alpha}^{*}$ in the CI scenario.}
 \label{algo:ci}
\end{algorithm}

\subsection{Inference under Partial Information}
\label{sec:methods_partial}

Once a degree of polynomial approximation $d$ is selected, the resulting approximation to the nonlinear term (Equation~\ref{eqn:Falpha}) can be substituted into the generalized UCM model (Equation~\ref{eqn:gen_UCM}) to define the TBF-CM as
\begin{equation}
\dot{\bm{\sigma}} - \bm{\nabla v}^{T} \cdot \bm{\sigma} - \bm{\sigma} \cdot \bm{\nabla v} + \dfrac{1}{\tau} \bm{\sigma} + \sum_{i=1}^{n_g} \sum_{j=1}^{n_b} \alpha_{i,j} B_j(\bm{\sigma}, \dot{\bm{\gamma}}) \, \bm{T}_i (\bm{\sigma}, \dot{\bm{\gamma}})= G  \dot{\bm{\gamma}}.
\label{eqn:rude2}
\end{equation}
The TBF-CM is a special case of the generalized UCM model like the Giesekus and PTT models. The nonlinear terms of the Giesekus and PTT models are parameterized by a single coefficient ($\alpha_G$ and $\epsilon_\text{PTT}$, respectively). On the other, the nonlinear term of the TBF-CM is parameterized by a set of coefficients $\bm{\alpha}$ that are likewise \textit{independent} of $\gamma_0$ and $\omega$. In other words, $\bm{\alpha}$ is a material property that is independent of the flow field. If $\bm{\alpha}$ is specified, then the TBF-CM is a fully parameterized nonlinear differential CM. This implies that we can use FLASH to solve for the PSS solution $\bm{\sigma}$ of the TBF-CM given $\bm{\alpha}$ and $\dot{\bm{\gamma}}$. FLASH relying on harmonic balance, internally transforms the resulting system of nonlinear differential equations into Fourier space and returns the Fourier coefficients $\hat{\bm{\sigma}}$ of the PSS solution truncated beyond the $(2H+1)$th harmonic. 

While it is easy to use inverse FFT to compute the time-domain representation, it is preferable to work in Fourier space for a few reasons. First, we can take Fourier transforms of any experimental data in the time domain to obtain $\hat{\bm{\sigma}}_\text{exp}$. A byproduct of this operation is that the high-frequency noise in $\bm{\sigma}_\text{exp}(t)$ is filtered. Second, the Fourier space provides a natural basis for describing periodic signals. A byproduct of this choice is that it allows us to compress data by exploiting the correlation between successive time-domain snapshots of $\bm{\sigma}$. 

In summary, given coefficients $\bm{\alpha}$ and the deformation gradient tensor $\dot{\bm{\gamma}}$ for OS flow, we can efficiently find the PSS $\hat{\bm{\sigma}}(\bm{\alpha}, \dot{\bm{\gamma}})$ by solving the TBF-CM using FLASH. Discovering a CM from experimental data corresponds to the inverse problem, i.e., inferring a sparse $\bm{\alpha}$ that yields $\hat{\bm{\sigma}}(\bm{\alpha}, \dot{\bm{\gamma}}) \approx \hat{\bm{\sigma}}_\text{exp}(\dot{\bm{\gamma}})$.




\subsubsection{Experimental Data and Loss Function}

Similarly to the CI scenario, we generate $n_\text{ds} = 12$ synthetic experimental datasets by using FLASH to solve the PTT model for $\bm{\sigma}_\text{exp}(t)$ (or $\hat{\bm{\sigma}}_\text{exp}$) in OS flow using $\epsilon_\text{PTT} = 0.3$ with $H = 8$ and $n_t = 2^6$. However, we ignore some of the components of the stress tensor in the PI scenario. We pretend that we only have access to shear stress $\sigma_{12}$ and discard information regarding normal stresses $\sigma_{11}$ and $\sigma_{22}$. 

Let $[\hat{\bm{\sigma}}^{(p)}_{12}]_\text{exp}$ denote the Fourier coefficients of the experimental PSS shear stress profile corresponding to the $p$th dataset ($1 \leq p \leq n_\text{ds}$) with the deformation gradient tensor $\dot{\bm{\gamma}}^{(p)}$ defined by the imposed strain frequency and amplitude. Only the subset $[\hat{\bm{\sigma}}^{(p)}_{12}]_\text{exp} \subset \hat{\bm{\sigma}}_\text{exp}^{(p)}$ is used for inference. 
Once we choose a degree of polynomial approximation $d$, the size of the coefficient vector $\bm{\alpha}$ is determined. Let $\hat{\bm{\sigma}}^{(p)}_{12}(\bm{\alpha}, \dot{\bm{\gamma}}^{(p)})$ denote the Fourier coefficients of the shear stress profile predicted by the TBF-CM with coefficients $\bm{\alpha}$ and deformation gradient tensor $\dot{\bm{\gamma}}^{(p)}$. 

We can evaluate how well a particular guess for $\bm{\alpha}$ matches experimental measurements by defining a LS loss function in Fourier space that sums over all the datasets,
\begin{equation}
\chi^{2}_{F}(\bm{\alpha}) = \dfrac{1}{2} \sum_{p=1}^{n_\text{ds}} \left\lVert [\hat{\bm{\sigma}}^{(p)}_{12}]_\text{exp} - \hat{\bm{\sigma}}_{12}^{(p)}(\bm{\alpha}, \dot{\bm{\gamma}}^{(p)}) \right\rVert^{2}_{2}.
\label{eqn:loss_chiF}
\end{equation}
Recall that both, $[\hat{\bm{\sigma}}^{(p)}_{12}]_\text{exp}$ and $\hat{\bm{\sigma}}^{(p)}_{12}(\bm{\alpha}, \dot{\bm{\gamma}}^{(p)})$, are vectors of the same size. It is possible to find an optimal $\bm{\alpha}$ by attempting to minimize $\chi^{2}_{F}(\bm{\alpha})$ using nonlinear LS. However, we do not pursue this direct approach because it does not impose sparsity on $\bm{\alpha}$, and is computationally challenging. Unlike the CI scenario, minimization of $\chi^{2}_{F}(\bm{\alpha})$ requires us to solve for $\hat{\bm{\sigma}}_{12}(\bm{\alpha}, \dot{\bm{\gamma}})$ under each of the $n_\text{ds}$ operating conditions during each iteration. If the cost per iteration were proportional to $n_\text{ds}$, it would not pose a major challenge for datasets of modest size, say $n_\text{ds} \sim \mathcal{O}(10) - \mathcal{O}(100)$ due to the speed of FLASH. However, if we use a minimizer that numerically computes the gradient $\nabla_{\bm{\alpha}} \chi_F^{2}$ to accelerate convergence,\cite{heath2018scientific} the computational cost rises sharply from $\mathcal{O}(n_\text{ds})$ to $\mathcal{O}(n_\text{ds} n_\alpha)$ per iteration, which can become exorbitant (see Section~\ref{sec:potential_comp_improvements} for a potential solution).

\subsubsection{Sparse Nonlinear Regression}

Recall that the SNLR problem, fashioned after Equation~\ref{eqn:sparse_nonlin_gen}, may be formulated as
\begin{equation}
\bm{\alpha}^{*} = \min_{\bm{\alpha}} \chi^{2}_{F}(\bm{\alpha}), \text{ such that } \Vert \bm{\alpha} \Vert_{0} \le k.
\label{eqn:opt_nonlin_sparse_cm}
\end{equation}

\begin{figure}
\begin{center}
\includegraphics[scale=0.7]{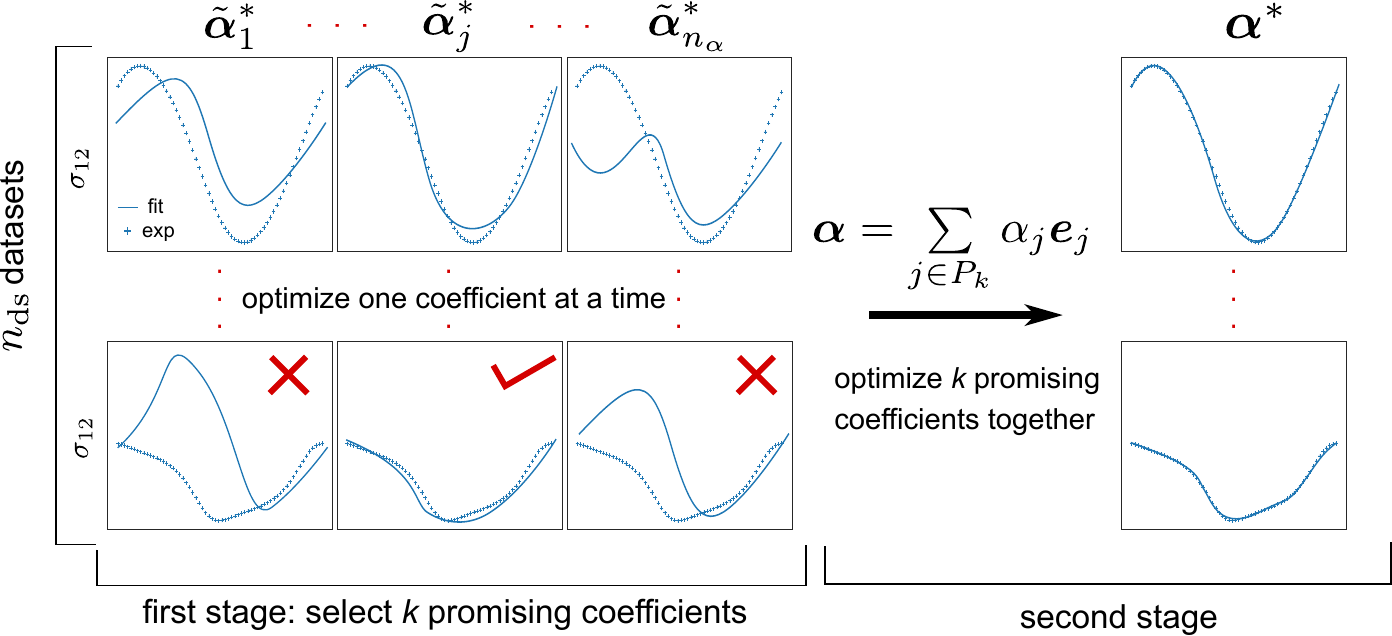}
\end{center}
\caption{Schematic of the greedy two-stage SNLR algorithm for the PI scenario. The first stage identifies the set $P_k$ of the $k$ most promising coefficients by performing $n_\alpha$ 1D optimizations. Sparsity is imposed by choosing $k \ll n_\alpha$. In the second stage, a $k$-dimensional optimization is performed by restricting the set of nonzero elements in $\bm{\alpha}$ to $P_k$. Although data and predictions are represented in the time domain, optimization is performed in the frequency domain in Algorithm 2.
 \label{fig:pi_schema}}
\end{figure}

We propose a simple greedy algorithm, schematically illustrated in Figure~\ref{fig:pi_schema}, to determine a sparse $\bm{\alpha}^{*}$. It is inspired by basis pursuit methods which attempt to identify and retain only the most promising features. We divide the task of inferring $\bm{\alpha}^{*}$ into two parts. In the first stage, we explore the potential of each individual element $\alpha_j \in \bm{\alpha}$, where $j \in [1, n_\alpha]$, to capture the data. In the second stage, we perform sparse nonlinear LS regression using only the $k$ most promising indices/coefficients. 

In the following, let $\bm{e}_{j}$ denote a unit vector of size $n_\alpha$ with a single nonzero element at the location $j \in [1, n_\alpha]$. In other words, the $j$th element of $\bm{e}_{j}$ is one, while the rest are equal to zero (similar to `one hot encoding' in ML). Then, $\tilde{\bm{\alpha}}_j = \alpha \bm{e}_j$ denotes a coefficient vector with a single nonzero element. In the first stage, we can perform a nonlinear LS minimization of $\chi^{2}_{F}(\tilde{\bm{\alpha}}_j)$ and determine the optimal value $\tilde{\bm{\alpha}}_j^{*}$, for each $j \in [1, n_\alpha]$. This is an independent 1D optimization for each coefficient that can be readily parallelized. For each 1D optimization, the cost per iteration is $\mathcal{O}(n_\text{ds})$ making the total cost $n_\alpha$ times the cost of each 1D optimization. Note that this is significantly better than $\mathcal{O}(n_\text{ds} n_\alpha)$ per iteration, since the cost \textit{per iteration} does not depend on $n_\alpha$, and the overall cost can be shared between multiple processors.

Promising coefficients are operationally defined as coefficients that result in small values of the loss function $\chi^{2}_{F}(\tilde{\bm{\alpha}}_j^{*})$. We sort $\chi^{2}_{F}(\tilde{\bm{\alpha}}_j^{*})$ for $j \in [1, n_\alpha]$ in ascending order and consider only the top $k \ll n_{\alpha}$ coefficients. Let $P_k$ denote the list of indices of these top $k$ most promising coefficients. The elements of $P_k$ are integers $j \in [1, n_\alpha]$ that mark the locations of these coefficients. The goal of the first stage is to furnish $P_k$.

In the second stage, we perform a nonlinear LS regression to minimize $\chi^{2}_{F}(\bm{\alpha})$, where sparsity is imposed on $\bm{\alpha}$ by construction. That is, we set
\begin{equation}
\bm{\alpha} = \sum_{j \in P_k} \alpha_j \bm{e}_j,
\label{eqn:constrained_alpha}
\end{equation}
where the summation runs over the list of $k$ elements in $P_k$. Thus, $\bm{\alpha}$ has only $k$ nonzero elements corresponding to the most promising indices. The cost per iteration at this stage is $\mathcal{O}(k n_\text{ds})$, where $k \ll n_\alpha$.

We solve all nonlinear minimization problems using the default trust-region reflective algorithm as implemented in the \texttt{scipy.optimize} library.\cite{Branch1999} For unbounded optimization, used in this work, this implementation is quite robust and similar to the implementation in MINPACK.\cite{More1978, More1980} The derivatives are computed numerically using a 2-point scheme. Algorithm \ref{algo:pi} summarizes the steps sketched in Figure~\ref{fig:pi_schema}. Input data consists of the Fourier coefficients of the shear stress and operating conditions $(\omega, \gamma_0)$ for each dataset. The degree of polynomial approximation $d$ and number of nonzero coefficients $k$ are also provided as input. Note that it is possible to determine a judicious value for $k \ll n_{\alpha}$ by examining the outcome of the first stage, in which the ability of each coefficient to describe experimental data is explored. The second stage performs a regular nonlinear LS calculation by allowing only the $k$ most promising coefficients to be nonzero.


\begin{algorithm}[h!]
\SetAlgoLined
\SetArgSty{textrm}
\DontPrintSemicolon
\KwIn{$\dot{\bm{\gamma}}^{(p)}$ and $[\hat{\bm{\sigma}}^{(p)}_{12}]_\text{exp}$  for $p = 1, \cdots, n_\text{ds}$; degree of approximation $d$; number of nonzero coefficients $k$.}
\KwResult{vector $\bm{\alpha}^{*}$ to approximate SCFs $\bm{g}(\bm{l})$}
\BlankLine
determine $n_\alpha$ from $d$ (Equation~\ref{eqn:glamBasis}).\;
\tcc{Stage 1: find promising coefficients using 1D optimization}
\For{$j \gets 1$ \KwTo $n_{\alpha}$}{
    let $\tilde{\bm{\alpha}}_j = \alpha \bm{e}_j$.\;
    minimize $\chi^{2}_{F}(\tilde{\bm{\alpha}}_j)$ and determine $\tilde{\bm{\alpha}}_j^{*}$ and  $\chi^{2}_{F}(\tilde{\bm{\alpha}}_j^{*})$.\;
}
argsort the loss functions $\{\chi^{2}_{F}(\tilde{\bm{\alpha}}_j^{*})\}_{j=1}^{n_\alpha}$ in ascending order\;
create list $P_k$ that marks the indices of the top $k$ most promising coefficients\;
\BlankLine
\tcc{Stage 2: find sparse solution with promising coefficients}
consider sparse $\bm{\alpha} = \sum_{j \in P_k} \alpha_j \bm{e}_j$ with $k$ nonzero elements\;
    minimize $\chi^{2}_{F}(\bm{\alpha})$ to determine $\bm{\alpha}^{*}$\;
 \caption{Greedy algorithm for SNLR to infer $\bm{\alpha}^{*}$ in the PI scenario.}
 \label{algo:pi}
\end{algorithm}

\section{Results}
\label{sec:results}

As mentioned previously, we generated $n_\text{ds} = 12$ synthetic datasets for the Giesekus ($\alpha_G = 0.3$) and PTT ($\epsilon_\text{PTT} = 0.3$) models and computed the PSS solution using FLASH with $H = 8$ and $n_t = 2^6$. We describe the results for the CI scenario, followed by the results for the PI scenario.

\begin{figure}
\begin{tabular}{cc}
\includegraphics[scale=0.45]{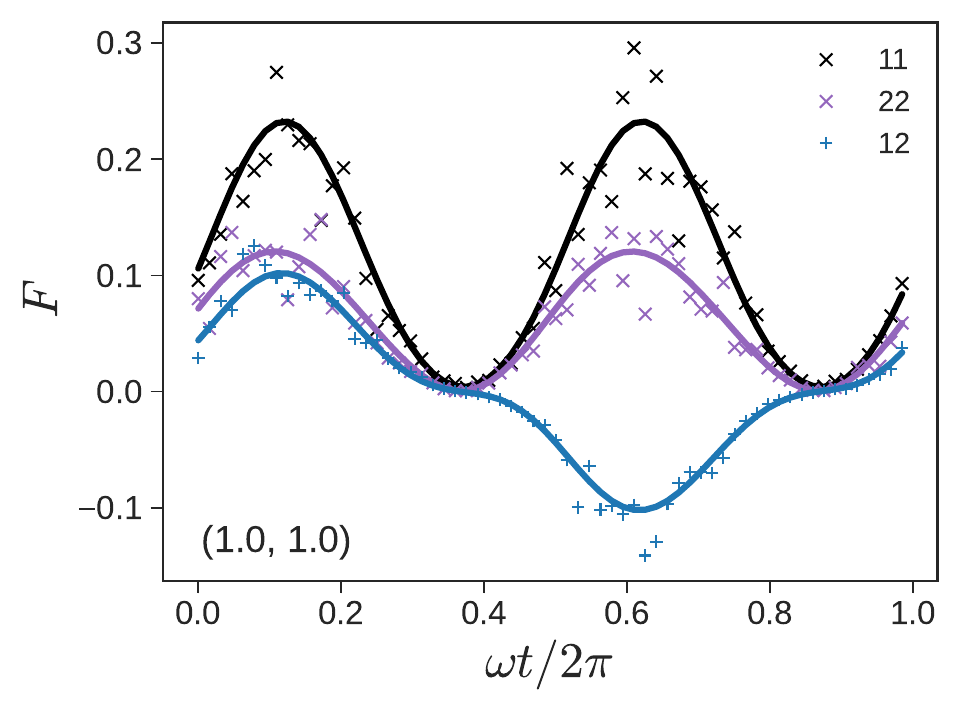} & \includegraphics[scale=0.45]{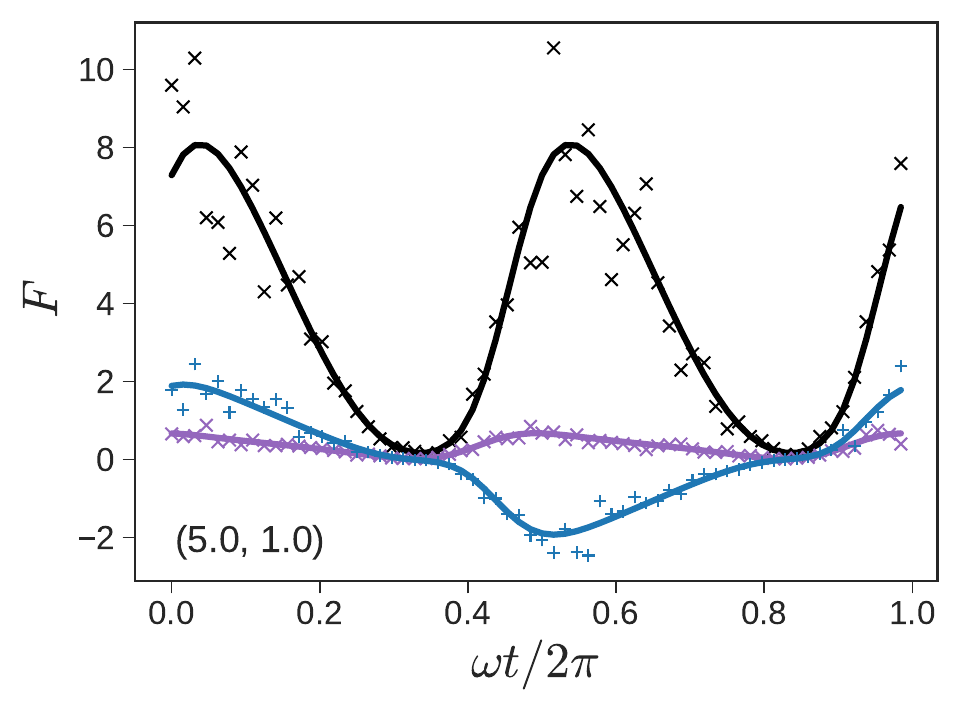} \\
(a) $\gamma_0 = 1, \omega \tau = 1$ & (b) $\gamma_0 = 5, \omega \tau = 1$
\end{tabular}
\caption{\textbf{Giesekus model in CI scenario}. Components of the nonlinear term $\bm{F}(t)$ (Equation~\ref{eqn:tbf_nonlin}) at $\omega \tau = 1$ and (a) $\gamma_0 = 1$ and (b) $\gamma_0 = 5$.  The crosses and plus signs show synthetic data, to which 20\% relative noise is added. Lines show fits obtained using SLR. \label{fig:giesekus_ftbf}}
\end{figure}

\subsection{Complete Information}
\label{sec:ci_results}

\subsubsection{Giesekus Model}

In the CI scenario, all components of the stress tensor are known, which allows us to evaluate the nonlinear term $\bm{F}(\bm{\sigma}, \dot{\bm{\gamma}})$. We added a relative Gaussian noise of 20\% to the nonlinear signal ($\bm{F}$) for the Giesekus model, as illustrated in Figure~\ref{fig:giesekus_ftbf} for two of the $n_\text{ds} = 12$ datasets. We perturbed $\bm{F}$ instead of $\bm{\sigma}$ because filtering noise from raw experimental data using standard Fourier analysis is a fairly common practice. We set the degree of polynomial approximation $d = 2$, which implies that the number of PFs is $n_b = 21$. The set of features includes one constant (1), five linear terms ($\{l_i\}_{i=1}^{5})$, and 15 unique quadratic ($\{l_i l_j\}_{i,j=1}^{5}$) terms. The number of unknowns (size of $\bm{\alpha}$) is $n_{\alpha} = n_g n_b = 189$. 

To contextualize the results of SLR, it is helpful to first consider the representation of the Giesekus model in terms of the TBFs. The form of the nonlinearity in the Giesekus model (Equation~\ref{eqn:F_giesekus}) coincides with one of the TBFs namely $\bm{T}_4 = \bm{\sigma} \cdot \bm{\sigma}$ (see Table \ref{tab:tbf_and_invariant}). Thus, $\bm{F}_\text{Giesekus} = \alpha_G/(G\tau) \bm{T}_4$, which implies that $g_4(\bm{l}) = \alpha_G/(G \tau)$ is a constant, while all other SCFs are zero, i.e., $g_i(\bm{l}) = 0$ for $i \ne 4$.

\begin{figure}
\centering
\begin{tabular}{cc}
\includegraphics[scale=0.4]{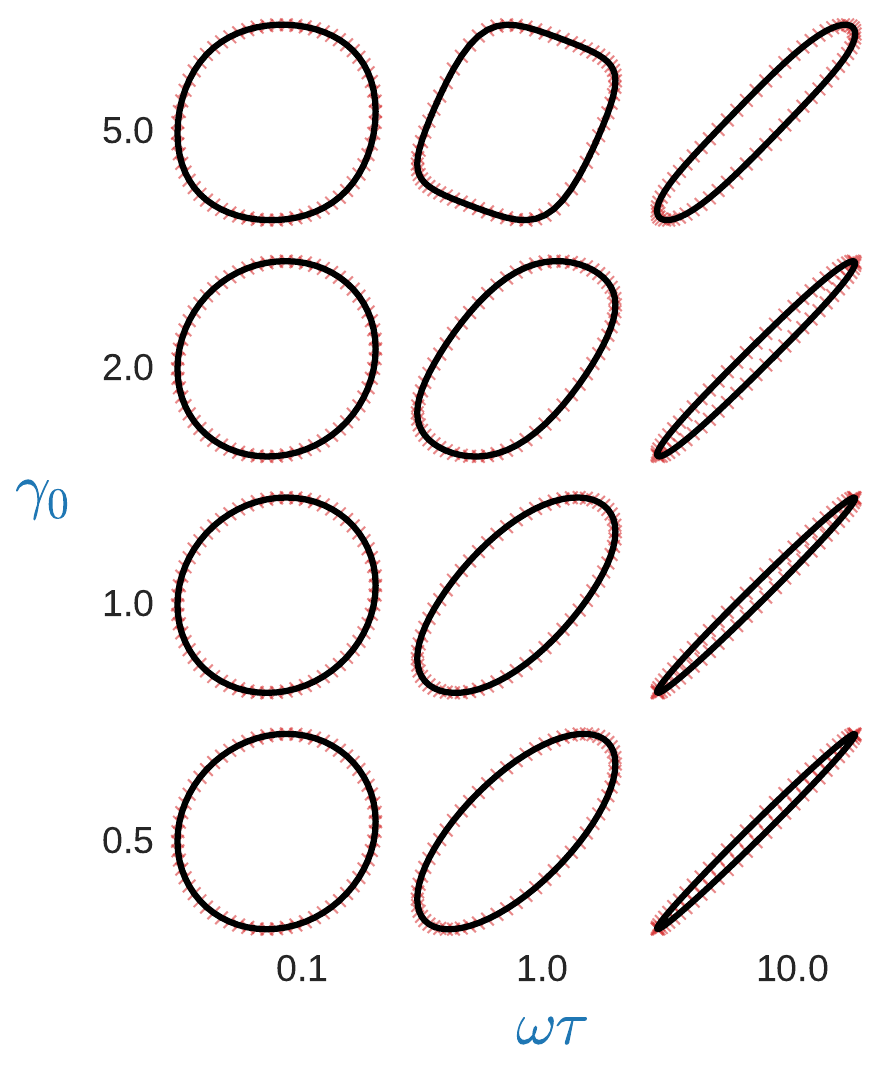} & 
\includegraphics[scale=0.4]{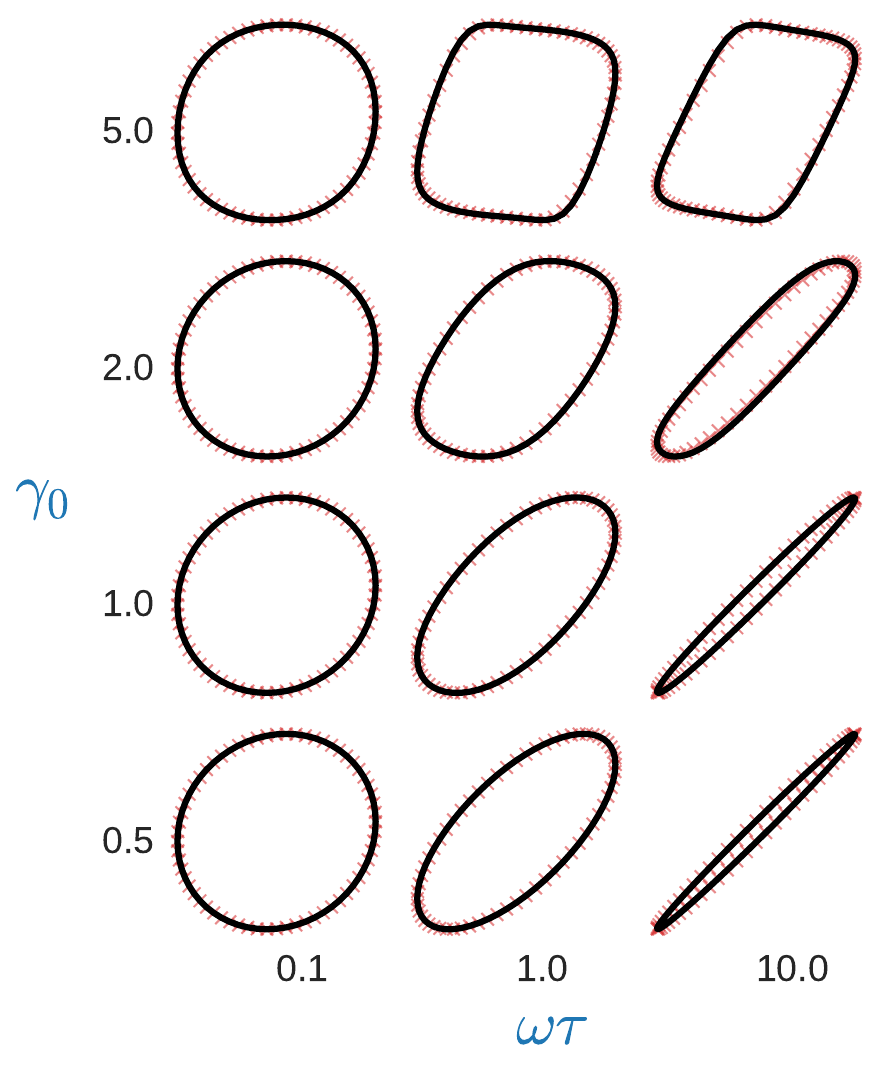} \\
(A) Giesekus & (B) PTT \\
\end{tabular}
\caption{Pipkin diagram showing the elastic Lissajous-Bowditch curves for (A) the Giesekus ($\alpha_G = 0.3$) and (B) PTT models. Similar to Figure~\ref{fig:giesekus_ftbf}, 20\% relative noise is added to $\bm{F}$. Symbols depict the normalized shear stress ($\sigma_{12}/\max(\sigma_{12})$) versus normalized strain ($\gamma/\gamma_0$), while lines depict the predictions of the CM discovered using SLR. \label{fig:giesekus_pipkin}}
\end{figure}

As expected, using Algorithm \ref{algo:ci}, we find that only one of these 189 coefficients (namely $\alpha_{4, 1}$) is nonzero. The regressed value of $\alpha_{4, 1} = 0.2988 \approx \alpha_G/(G\tau)$ shows that this technique identifies the true CM, at least for this simple example. The coefficient of correlation $R^{2} = 0.97$. This example serves as a validation test of the proposed method. Furthermore, this calculation is fast, since the problem is linear. It took about 0.5  sec on an old commodity desktop computer (Intel i7-6700 3.4 GHz CPU). Recall that the CI scenario, unlike the PI scenario, does not require the computation of the PSS solution using FLASH or numerical integration to find the optimal set of coefficients.

An elastic Lissajous-Bowditch curve is a parametric plot of stress versus strain. Figure~\ref{fig:giesekus_pipkin} depicts a Pipkin diagram which is a table of normalized elastic Lissajous-Bowditch curves for the $n_\text{ds}$ = 12 datasets. Here, the shear stress ($\sigma_{12}$ component of $\bm{\sigma}$), normalized by its maximum value, is plotted against the normalized strain $\gamma(t)/\gamma_0 = \sin \omega t$. Symbols denote synthetic data (20\% relative noise is only added to $\bm{F}$), while lines show the TBF-CM which is nearly identical to the true CM in this case.

In a typical use case, $\bm{F}$ is not noisy since the measured signal ($\bm{\sigma}$) can be filtered using Fourier analysis. If we use a noiseless $\bm{F}$ as input, then the true CM $\alpha_{4, 1} = 0.3 = \alpha_G/(G \tau)$ is discovered using Algorithm \ref{algo:ci}. The true CM with a single nonzero component is discovered even when we increase $d = 3$ corresponding to $n_\alpha = 506$, or $d = 4$ corresponding to $n_\alpha = 1134$. As expected, the computational time increases with $d$ as the size of the regression problem increases. The CPU time increases from $\sim$ 0.5 s for $d = 2$ to $\sim$ 1 s for $d = 3$ and $\sim$ 3 s for $d = 4$.

Thus, we validated Algorithm \ref{algo:ci} for the Giesekus model and found that it is able to discover the true CM in the CI scenario using only the PSS stress profile for datasets of modest size ($n_\text{ds} = 12$). Indeed, for the Giesekus model, which has a particularly simple representation in terms of the TBFs, we found that synthetic data generated at a single value of $\gamma_0 \gtrsim 2$ (that is, $n_\text{ds}$ = 3, due to three frequencies) were sufficient to discover the true CM. The value of $\gamma_0$ must be large enough to trigger a detectable nonlinear response, otherwise $\bm{F} \approx \bm{0}$ does not contain enough information. SLR is remarkably fast and fairly robust (it is not particularly sensitive to perturbations $\sim$20\% in $\bm{F}$), because the optimization problem is well posed.

While this example provides a glimpse of the promise of this method, caution is warranted. First, the CI scenario is not representative of practical problems. Second, the Giesekus model plays into the strengths of the method because of its simple structure. Therefore, we consider the PTT model under the CI scenario next, before moving to the harder problem of PI.

\subsubsection{PTT Model}

As before, we generated $n_\text{ds} = 12$ synthetic datasets using the exponential PTT model with parameter $\epsilon_\text{PTT} = 0.3$. We set $d = 2$ and obtained the nonlinear term $\bm{F}(\bm{\sigma}, \dot{\bm{\gamma}})$. Of the $n_\alpha = 189$ possible coefficients, Algorithm \ref{algo:ci} identified only two nonzero components of $\bm{\alpha}$. Both these coefficients, namely $\alpha_{2, 2} = 0.306$ and $\alpha_{2, 5} = 0.020$, correspond to the SCF $g_2$. All other SCFs were zero. Therefore, according to SLR, the nonlinear term corresponding to data generated from the PTT model is,
\begin{equation}
\bm{F}_\text{PTT}^{\text{SLR}} = g_2 \bm{T}_2 = \left( 0.306\, I_1 + 0.020\, I_4 \right) \bm{\sigma},
\label{eqn:Fptt_lasso}
\end{equation}
where $I_1 = \text{tr}(\bm{\sigma})$ and $I_4 =  \text{tr}(\bm{\sigma} \cdot \bm{\sigma} \cdot \bm{\sigma})$. The Pipkin diagram comparing the experimental elastic Lissajous-Bowditch curves with the inferred CM is shown in Figure~\ref{fig:giesekus_pipkin}B.

Unlike the Giesekus model, the exponential PTT model cannot be easily decomposed in terms of TBFs, except in special situations. In the limit of small $\epsilon_\text{PTT}$ and $I_1/G$, the nonlinear term can be approximated as
\begin{equation}
\bm{F}_\text{PTT} = \dfrac{\exp\left(\dfrac{\epsilon_\text{PTT}}{G} I_1 \right) - 1}{\tau} \bm{\sigma} 
\approx \left[ \dfrac{\epsilon_\text{PTT}}{G\tau} I_1 +  \dfrac{1}{2} \dfrac{\epsilon^{2}_\text{PTT} I_1^{2}}{G^2 \tau} \right] \bm{T}_2.
\label{eqn:ptt_tbf_taylor}
\end{equation}
Terms of order $(\epsilon_\text{PTT} I_1/G)^3$ and higher are neglected in the Taylor series expansion in Equation~\ref{eqn:ptt_tbf_taylor}. Thus, the only nonzero SCF is $g_2$, which only depends on $I_1$. Comparison of $g_2$ in eqns  \ref{eqn:Fptt_lasso} and \ref{eqn:ptt_tbf_taylor} indicates that the prefactor to $I_1$ (0.306) is comparable with $\epsilon_\text{PTT}/(G \tau) = 0.3$. However, the second term in Equation~\ref{eqn:Fptt_lasso}, $0.020\, I_4$, is different from the $0.045 I_1^2$, expected from Equation~\ref{eqn:ptt_tbf_taylor}. Interestingly, if we use $\epsilon_\text{PTT} = $ 0.1 instead of 0.3 to generate the synthetic datasets, SLR identifies a single nonzero coefficient and $\bm{F}_\text{PTT}^{\text{LASSO}} = 0.309 I_1 \bm{\sigma}$, which is similar to Equation~\ref{eqn:ptt_tbf_taylor}.

As in the Giesekus model, increasing the polynomial approximation to $d = 4$ effectively discovers the same TBF-CM, but takes $\sim$ 2.5 s instead of $\sim$ 0.5 s for $d = 2$. Unlike the Giesekus model, a single LAOS dataset is no longer sufficient to infer the CM. With some experimentation, we found that the CM with a nonlinear term given by Equation~\ref{eqn:Fptt_lasso} is usually discovered, even if we use only three values of $\gamma_0$ instead of four, e.g. $\gamma_0 = \{0.5, 1.0, 2.0\}$. That is, $n_\text{ds} = 9$ appears to be sufficient for the PTT model. The key generalizable lesson from these examples for experimental design is to collect sufficient data in the LAOS regime (roughly, $\gamma_0 \gtrsim 1$).

\subsection{Partial Information}
\label{sec:pi_results}
 In this section, we demonstrate the application of the SNLR algorithm (Algorithm \ref{algo:pi}) to synthetic data obtained using the PTT model with $\epsilon_\text{PTT} = 0.3$ under $n_\text{ds} = 12$ different operating conditions. However, unlike the CI scenario, we only use shear stress data for inference. We focus on the PTT model because it produces more interesting results and offers generalizable lessons that are relevant for CM discovery. Results for the Giesekus model are relegated to supplementary material Section~\ref{sec:Giesekus_PI}.

\begin{figure}
\centering
\includegraphics[scale=0.7]{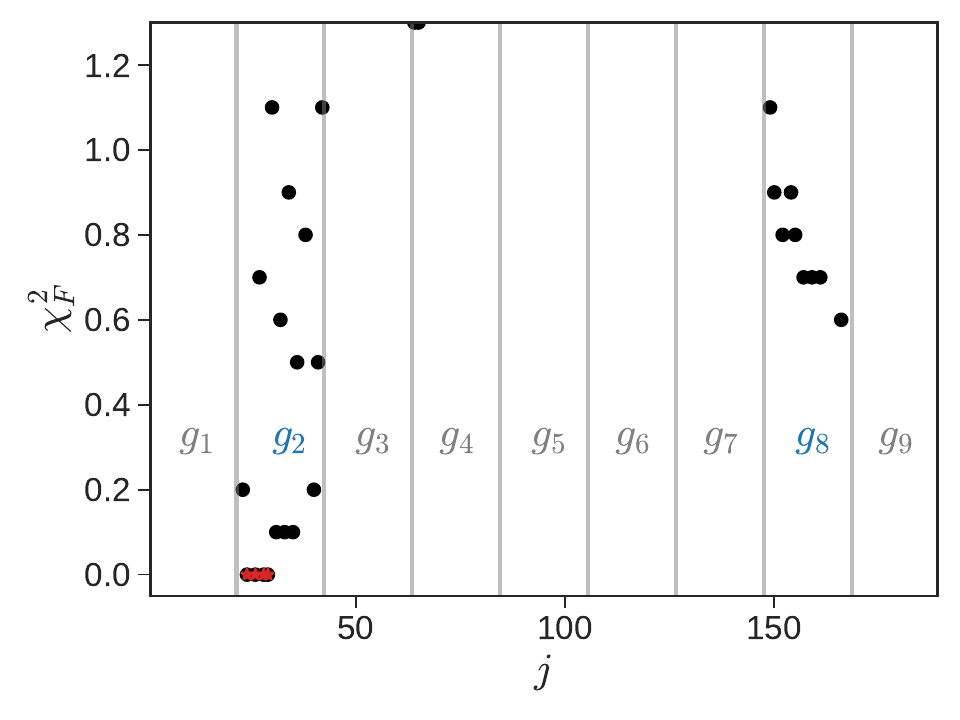}
\caption{Loss functions $\chi^2_F(\tilde{\bm{\alpha}}_j)$, where $j \in [1, n_\alpha = 189]$, obtained after the first stage of Algorithm \ref{algo:pi} with $d=2$. The gray vertical lines demarcate the coefficients that belong to different SCFs. The upper limit of the graph is truncated to emphasize the spread among promising coefficients. The cluster of red stars corresponds to the four smallest values of $\chi^2_F(\tilde{\bm{\alpha}}_j)$, and highlights the $k=4$ most promising coefficients. \label{fig:optim1c}}
\end{figure}

As before, we set $d = 2$, which implies $n_{\alpha} = 189$. In the first stage, we scan through $\tilde{\bm{\alpha}}_j$ for $j \in [1, n_{\alpha} = 189]$ by performing $n_\alpha$ independent 1D nonlinear minimizations to obtain optimal values $\tilde{\bm{\alpha}}_j^{*}$ and the corresponding loss functions $\chi^{2}_{F}(\tilde{\bm{\alpha}}_j^{*})$. The average cost of each of these minimizations was 25.2 $\pm$ 16.4 s, or approximately half a minute. On a computer with a single core, the first stage can be completed in about an hour and a half. On a workstation with 16 cores, this cost reduces to a little under 6 minutes, due to the ease of parallelization. 

The computational cost is proportional $n_{\alpha}$, which implies that the cost of the first stage increases rapidly with $d$. If we increase the degree of approximation from $d = 2$ to $d = 4$, the total computational cost increases 6x, assuming that the average cost per minimization remains unchanged. On a workstation with 16 cores, the total runtime for the first stage increases to about 40 minutes. 

The $\chi^{2}_{F}$ corresponding to the most promising coefficients is shown in Figure~\ref{fig:optim1c}. The first stage correctly identifies the importance of $g_2$ for the PTT model. The top $k = 4$ most promising coefficients form a distinct cluster with $\chi^{2}_{F} < 0.1$, and are indicated by red stars. In order of increasing $\chi^{2}_{F}$, these coefficients, namely $\alpha_{2, 7} \equiv \alpha_{28}$, $\alpha_{2, 3} \equiv \alpha_{24}$, $\alpha_{2, 8} \equiv \alpha_{29}$, and $\alpha_{2, 5} \equiv \alpha_{26}$, correspond to PFs $I_1^2$, $I_2$, $I_1 I_2$, and $I_4$, respectively.

\begin{figure}
\centering
\includegraphics[scale=0.5]{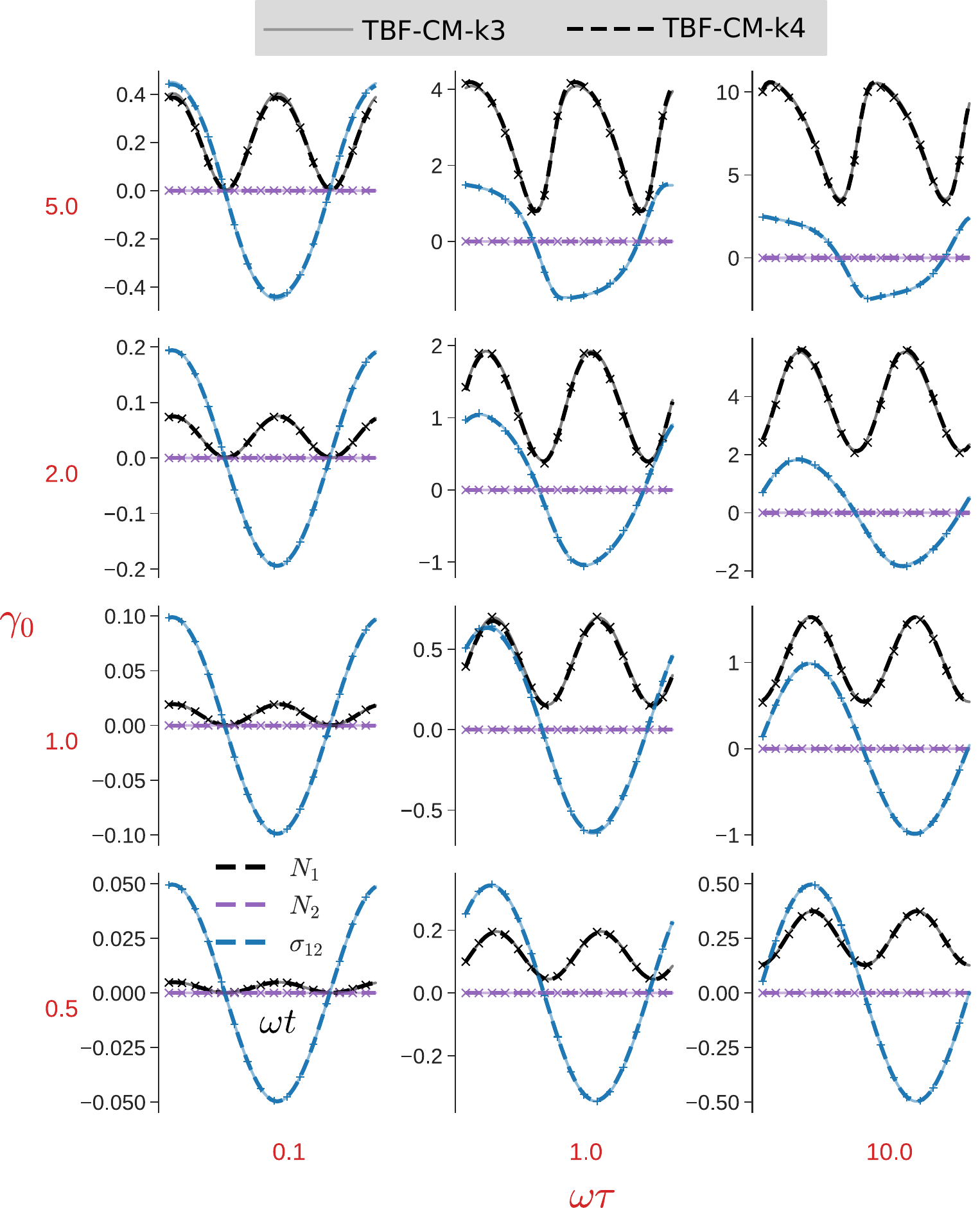}
\caption{\textbf{PI LAOS Training}. Fits of two different TBF-CMs (dashed and light solid lines correspond to TBF-CM-k4 and TBM-CM-k3 respectively; see text for details) to synthetic data generated from the PTT model with $\epsilon_\text{PTT} = 0.3$ (symbols). Only shear stress data (blue) are used to infer the TBF-CMs, but predictions of $N_1$ (black) and $N_2$ (purple) are also included for comparison. \label{fig:fitTBF_OS}}
\end{figure}

For the second stage of Algorithm \ref{algo:pi}, in addition to $k = 4$, we also entertained the idea of retaining only the top $k = 3$ coefficients to explore the possibility of multiple solutions, and sensitivity to the choice of $k$. Thus, we considered two different sets of nonzero coefficients, $P_{4} = [28, 24, 29, 26]$, and $P_{3} = [28, 24, 29]$. That is, in the set $P_{4}$, we picked the top $k=4$ most promising coefficients, while in the set $P_{3}$, we selected only the top $k=3$ most promising coefficients. We then performed a nonlinear regression that allowed only these coefficients to vary and ended up with two distinct TBF-CMs. They correspond to,
\begin{align}
\bm{F}_{k4}^\text{SNLR} & =  \left(0.3375 I_2 - 0.1637 I_4 - 0.2770\, I_1^2 + 0.1794\, I_1 I_2 \right) \bm{T}_2 \notag \\
\bm{F}_{k3}^\text{SNLR} & =  \left(0.2461 I_2 - 0.2174\, I_1^2  + 0.0114\, I_1 I_2 \right) \bm{T}_2.
\label{eqn:top3tbf}
\end{align} 
Hereafter, we label these two TBF-CMs as TBF-CM-k4 and TBF-CM-k3. The computational cost of the second stage of the SNLR algorithm is relatively small compared to that of the first stage. For instance, TBF-CM-k4 required 8 iterations which took $\sim$40 sec, while TBF-CM-k3 required only 6 iterations, which took $\sim$25 s. The final values of the loss function $\chi^{2}_{F}$ after optimization were $1.39 \times 10^{-3}$ and $3.04 \times 10^{-3}$ for TBF-CM-k4 and TBF-CM-k3, respectively.

The fits of these two TBF-CMs and the experimental data are shown in Figure~\ref{fig:fitTBF_OS}, where the dashed and thin lines represent TBF-CM-k4 and TBF-CM-k3, respectively. The symbols show the output of the PTT model. Although $\sigma_{12}$, $N_1$ and $N_2$ are all shown, it is important to remember that \textit{only the shear stress component is used}  in both stages of Algorithm \ref{algo:pi}. Visually, the agreement of both TBF-CMs with the data is excellent across the board. Experiments with $k = 1$ and $k = 2$, produced fits that were qualitatively similar, but visually inferior.

This success is noteworthy for at least two reasons. First, although only shear stress data is used for inference in the PI scenario, the agreement of the predicted $N_1$ and $N_2$ with the PTT model is remarkable. Second, TBF-CM-k4 and TBF-CM-k3 look quite different from the TBF-CM inferred in the CI scenario, namely $\left( 0.306 I_1 + 0.020 I_4 \right) \bm{T}_2$, which we subsequently label TBF-CM-CI, and the Taylor expansion of the PTT model in the limit of small $\epsilon_\text{PTT}$ and $I_1/G$, namely $\left( 0.3 I_1 + 0.045 I_1^2 \right) \bm{T}_2$. We examine the origin and implication of these two observations in succession.

Although the agreement of $N_1$ and $N_2$ predicted by the TBF-CMs with the PTT model is a fortunate accident that is unlikely to generalize to real materials, it is aided by the strict enforcement of physical constraints, and bias for parsimony.  In particular, we observe that both TBF-CM-k4 and TBF-CM-k3 correctly predict $N_2 = 0$. Retrospectively, this can be interpreted as a direct consequence of $\bm{F}(\bm{\sigma}, \dot{\bm{\gamma}}) = g_2 \bm{T}_2 = g_2 \bm{\sigma}$. When the nonlinear term is proportional to $\bm{\sigma}$, $\sigma_{22}(t) = 0$ remains unperturbed, if the initial condition $\bm{\sigma}(0) = \bm{0}$, regardless of the functional form of $g_2(\bm{l})$.

The extraordinary agreement of $N_1$ and $N_2$ predicted by the TBF-CMs and the unseen synthetic data can be traced to the common origin of the nonlinear term for $\sigma_{11}$, $\sigma_{22}$, and $\sigma_{12}$. In particular, the components of $\bm{F}_\text{PTT}(\bm{\sigma}, \dot{\bm{\gamma}})$ for the PTT model in Equation~\ref{eqn:F_PTT} corresponding to these terms take the form $F_{11} = g_2(\bm{l}) \sigma_{11}$, $F_{22} = g_2(\bm{l}) \sigma_{22}$, and $F_{12}(\bm{l}) = g_2(\bm{l}) \sigma_{12}$. Thus, if a good approximation for $g_2(\bm{l})$ is obtained from the shear stress data, we expect it to also be a reasonable approximation for $\sigma_{11}$ and $\sigma_{22}$. 

What about the uniqueness of the inferred TBF-CM? TBF-CM-k4 and TBF-CM-k3 are products of Algorithm \ref{algo:pi}, which is a crude method for SNLR. In the PI scenario, we \textit{expect} different CMs to result from different algorithms and parameter choices. The fact that at least three different CMs (TBF-CM-k4, TBF-CM-k3, and TBF-CM-CI) successfully describe shear stress data from a simple model like the PTT model underscores the inherent nonuniqueness of any CM discovered from limited data. Somewhat paradoxically, this nonuniqueness highlights the importance of frame invariance and inductive biases in navigating the space of potential CMs. 

\subsubsection{Generalizability of TBF-CMs}

The TBF-CMs inferred using the proposed algorithms fit  experimental or training datasets remarkably well. However, what we usually care about for CFD applications is the ability of the learned model to generalize beyond the data used to train it. In this work, we know the true CMs (Giesekus or PTT). Therefore, we can compare the predictions of the inferred TBF-CMs with the true CMs beyond the training data.

We assess the generalizability of TBF-CM-k4 and TBF-CM-k3 by performing two different \textit{types} of tests. In the first type, we consider the predictions of the TBF-CMs in OS flow, where the operating conditions are different from those used in the $n_\text{ds}$ training datasets. Some of the $(\gamma_0, \omega)$ in these tests lie within the region circumscribed by the experimental data. We label these predictions as \textit{interpolations}. In other cases,  $(\gamma_0, \omega)$ lie outside the range of the training data. We label these predictions of the inferred TBF-CMs as \textit{extrapolations}. 

\begin{figure}
\centering
\includegraphics[scale=0.4]{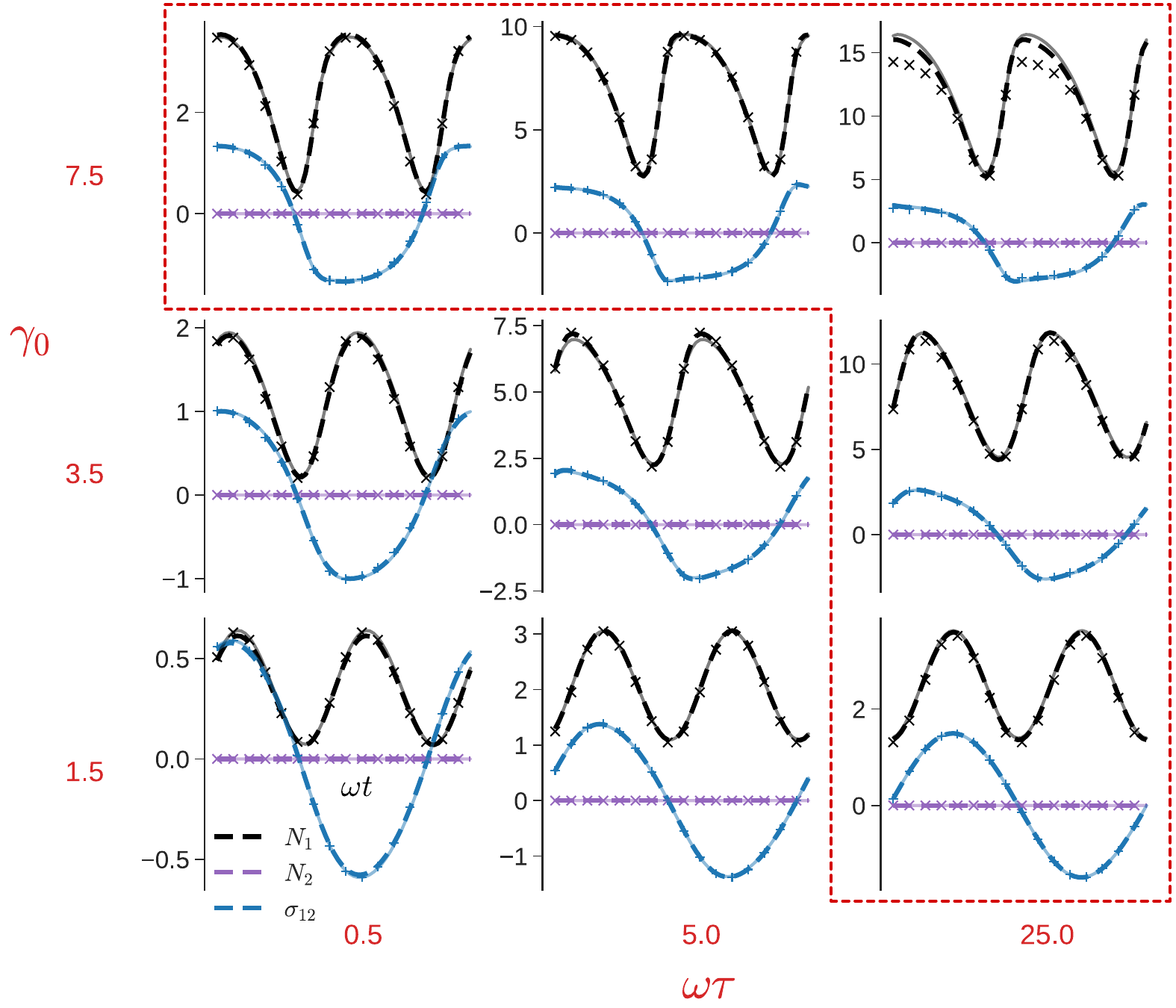}
\caption{\textbf{PI OS flow Test}. Interpolation (region outside the dashed line) and extrapolation (region inside dashed line) predictions of TBF-CM-k4 (dashed lines) and TBF-CM-k3 (light solid line) are compared with the response of the PTT model with $\epsilon_\text{PTT} = 0.3$ (symbols). Each subplot shows the PSS solution for $N_1$ (black), $N_2$ (purple), and $\sigma_{12}$ (blue) as a function of $\omega t \in [0, 2\pi]$.
 \label{fig:predictTBF_OS}}
\end{figure}

In the second type of experiment, we test the ability of the inferred CMs to generalize beyond the OS flow. These are much tougher benchmarks. In particular, we consider the predictions of the TBF-CMs to the startup of steady shear flow in which a constant shear rate $\dot{\gamma}$ is applied to an initially relaxed material ($\bm{\sigma}(0) = \bm{0}$). Finally, we transition from shear flows to uniaxial extensional flows in which we consider the evolution of the normal stresses when a steady elongational strain rate $\dot{\epsilon}$ is applied.\cite{morrison2001understanding}

In Figure~\ref{fig:predictTBF_OS}, we show the results of the first type of test (OS shear). Recall from Figure~\ref{fig:fitTBF_OS}, that the range of training data span $0.5 \leq \gamma_0 \leq 5$, and $0.1 \leq \omega \tau \leq 10$. The four tests in the lower left corner of Figure~\ref{fig:predictTBF_OS} with $(\gamma_0, \omega \tau)$ equal to $(1.5, 0.5)$, $(1.5, 5.0)$, $(3.5, 0.5)$, and $(3.5, 5.0)$ lie within the convex hull of the training data. Thus, they represent interpolations. The other five datasets in the figure marked off by the red dashed line lie outside the range of training data and hence represent extrapolations. 

Visually, we find that the predictions of TBF-CM-k4 and TBF-CM-k3 are comparable and largely agree with the PTT model. On closer inspection, we find that TBF-CM-k4 is slightly superior, especially in describing $N_1$. As $\gamma_0$ increases well beyond the range of the training data (top right corner of Figure~\ref{fig:predictTBF_OS}),  predictions of $N_1$ start to become less satisfactory. From numerical experiments, we find that both the TBF-CMs extrapolate satisfactorily beyond the training data  (see supplementary material Section~\ref{sec:extrap_pi_os}). At large $\omega \tau \gtrsim 50$ and strain amplitude ($\gamma_0 \gtrsim 7.5$), it is advisable to resolve a greater number of harmonics (use larger $H > 8$), and/or use more intermediate steps to aid the convergence of FLASH. Regardless, TBF-CMs using Algorithm \ref{algo:pi} appear to generalize well for the PTT model in OS flow.


Next, we consider the second, more stringent, type of test in which we probe the behavior of TBF-CM-k4 and TBF-CM-k3 when a constant deformation rate is applied under shear and uniaxial extension.  In steady shear flow, the velocity gradient tensor $\bm{\nabla v} = \dot{\gamma} \bm{e}_{12}$ and the deformation gradient tensor $\dot{\bm{\gamma}} = \dot{\gamma} \bm{e}_{12} + \dot{\gamma} \bm{e}_{21}$, are the same as for OS flow. The only difference is that the shear rate $\dot{\gamma}$ is constant rather than sinusoidal. In uniaxial extension, the velocity gradient tensor is diagonal $\bm{\nabla v} = \dot{\epsilon}\, \bm{e}_{11} - 0.5 \dot{\epsilon}\, \bm{e}_{22} - 0.5 \dot{\epsilon}\, \bm{e}_{33}$, where $\dot{\epsilon}$ is the steady extensional strain rate. This implies that the shear rate tensor $\dot{\bm{\gamma}} = \bm{\nabla v} + (\bm{\nabla v})^{T} = 2 \bm{\nabla v}$ and the stress tensor $\bm{\sigma} = \sigma_{11}  \bm{e}_{11} + \sigma_{22}  \bm{e}_{22} + \sigma_{33}  \bm{e}_{33}$ have three nonzero components along the diagonal. TBFs and invariants specialized for uniaxial extension are presented in the supplementary material (Section~\ref{sec:TBF_unaxial}).

\begin{figure}
\centering
\includegraphics[scale=0.5]{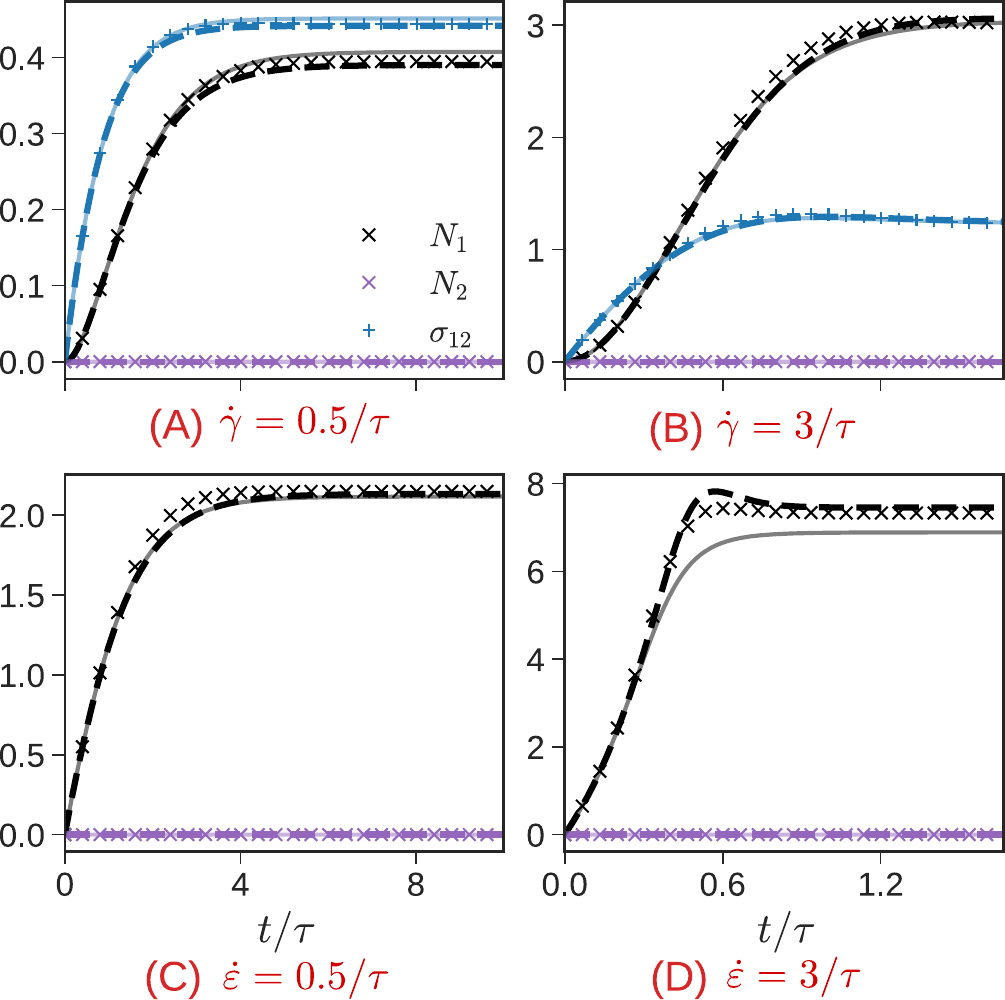}
\caption{\textbf{PI startup shear and extension}. Startup of shear (A and B) and unaxial extension (C and D) for the PTT model (depicted by symbols) at two different deformation rates. The predictions of TBF-CM-k4 and TBF-CM-k3 are shown by the dashed and light solid lines, respectively.\label{fig:startup}}
\end{figure}

We solve the initial value problem using an implicit Runge-Kutta method belonging to the Radau IIA family implemented in the Python package \texttt{scipy}.\cite{Virtanen2020, Hairer1996} It is a fifth-order method with local error controlled via a third-order accurate embedded formula. We assume that stress tensor components  are initially zero, i.e., $\bm{\sigma}(0) = \bm{0}$. In steady shear, we compute the evolution of the shear and normal stress components, whereas in steady uniaxial extension, we compute the evolution of only the normal stress components, since $\sigma_{12} = 0$. For the PTT model, $N_2 = 0$, as mentioned earlier.

We consider two different deformation rates ($\dot{\gamma}$ and $\dot{\epsilon}$) of 0.5/$\tau$ and 3/$\tau$. Comparisons of the PTT model and the TBF-CMs are shown in Figure~\ref{fig:startup}. For the startup of steady shear, both TBF-CMs appear to work reasonably well. TBF-CM-k4 appears to be slightly more accurate in predicting shear stress. For $N_1$, both models appear to capture the initial growth, but slightly under- or over-estimate the steady-state value. 

A similar trend is also observed in the uniaxial extension. TBF-CM-k4 describes $N_1$ better, especially at $\dot{\epsilon} = 3/\tau$, where TBF-CM-k3 under-estimates the steady state. It captures the mild overshoot in $N_1$ at large $\dot{\epsilon}$ semi-quantitatively, but exaggerates the bump. Nevertheless, these out-of-sample tests show that the TBF-CMs inferred from PI using only $\sigma_{12}$ data in OS experiments, generalize surprisingly well to startup shear and extension. Therefore, we speculate that they will be reasonable approximations to the PTT model in CFD simulations of non-homogeneous flows encountered during processing.

\section{Discussion}
\label{sec:discussion}

Although the proposed algorithms are presented only as proof-of-concept ideas in this work, they appear to be promising tools for CM discovery from experimental data. They have important implications for experimental design and algorithmic improvement.

\subsection{Implications for Design of Experiments}
\label{sec:exp_design}

First, it appears that modest datasets with $n_\text{ds} \sim \mathcal{O}(10)$, typical in rheological settings, may be sufficient to infer CMs that obey physical constraints and extrapolate reasonably well beyond the scope of the training data. Second, in OS experiments where the shear stress and both normal stress differences are measured, we can use SLR, which is fast and robust. Unfortunately, the cost of posing a computationally desirable problem involves shifting the burden from the modeler to the experimenter. Third, OS experiments \textit{need} to probe strongly nonlinear flows by exploring the large $\gamma_0$ regime to induce a sufficiently robust nonlinear signal. In this work, we weighted the $n_\text{ds}$ datasets equally. This potentially runs the risk of diluting more valuable information from measurements at large $\gamma_0$. We plan to explore this issue further in future work. In terms of frequency, we recommend exploring both the $\omega \tau < 1$ and $\omega \tau > 1$ regimes.

A deeper lesson arises from the fact that only five of the nine simultaneous invariants of $\bm{\sigma}$ and $\dot{\bm{\gamma}}$ are activated in shear flow. This places theoretical limits on the types of CMs that can be discovered using only shear flows, even in the CI scenario. In other words, it is possible to find TBF-CMs that perfectly describe the response to all shear flows, but fail to describe material behavior when the flow field includes extensional components. An important practical outcome of this insight is the necessity of combining shear and extension data for the CM discovery of real materials.

\subsection{Potential for Computational Improvements}
\label{sec:potential_comp_improvements}

Before we discuss potential improvements, it is helpful to emphasize that the TBF framework is only guaranteed to work when the nonlinear term $\bm{F}$ in the CM is analytical. This is not the case for some CMs of transient networks,\cite{Ahn1995, Mittal2024a} polymers,\cite{Ianniruberto2001} and yield-stress materials,\cite{Larson1988, morrison2001understanding} which contain singularities. It is worthwhile to empirically test how the TBF-CM framework fares for such CMs. Futhermore, if the nonlinear term $\bm{F}(\bm{\sigma}, \dot{\bm{\gamma}})$ depends on terms other than $\bm{\sigma}$ and  $\dot{\bm{\gamma}}$, then the set of TBFs has to appropriately modified.

Assuming $\bm{F}(\bm{\sigma}, \dot{\bm{\gamma}})$ is analytical, a simple improvement of the computational protocol is the relaxation of using a fixed $H$ for all experimental datasets. We kept $H = 8$ fixed in this work to simplify exposition. In practice, we can use Fourier analysis of experimental data to calibrate the appropriate $H$ for each dataset. This can potentially speed up the calculation in the PI scenario.

Algorithm \ref{algo:pi} proposed for the PI scenario is arguably too simple. While it works well with simple CMs like the Giesekus and PTT models, its suitability for more complex materials remains to be explored. The most obvious improvement is to use better algorithms for SNLR, informed by the rapid theoretical and algorithmic progress in nonlinear compressed sensing over the past decade or so. Most of the proposed algorithms such as iterative hard thresholding,\cite{Blumensath2008a} greedy sparse simplex,\cite{Beck2013} gradient pursuit,\cite{Blumensath2008} etc. involve taking the gradient of the loss function $\nabla_{\bm{\alpha}} \chi^{2}_F$ at each iteration. A two-point numerical computation of $\partial{\hat{\bm{\sigma}}}/\partial \alpha_j$ required to evaluate this gradient would involve $\mathcal{O}(n_\alpha)$ FLASH simulations \textit{per iteration}. This is prohibitively expensive. Fortunately, the TBF-CM is linear in the coeffficient vector $\bm{\alpha}$. Consequently, terms such as $\partial{\hat{\bm{\sigma}}}/\partial \alpha_i$ can be computed using a combination of FFT and inverse FFT.\cite{Krack2019} This means that a single FLASH simulation per iteration is sufficient to assemble $\nabla_{\bm{\alpha}} \chi^{2}_F$. The asymptotic computational cost of this assembly scales as $\mathcal{O}(n_\alpha)$ because $\nabla_{\bm{\alpha}} \chi^{2}_F$ has $n_\alpha$ components. However, computing each component ($\partial{\hat{\bm{\sigma}}}/\partial \alpha_i$) is relatively cheap since it only involves an FFT and its inverse, which costs $\mathcal{O}(n_t \log n_t)$. 

Finally, we have to tackle the problem of abundance. Depending on $k$ (which can be determined using clustering methods) and other details of the SNLR algorithm, we usually obtain more than one potential TBF-CM that fits the data. One possibility to deal with such non-uniqueness is to reformulate the inverse problem as a sampling problem using a Bayesian formulation, where a distribution of CMs is entertained,\cite{Shanbhag2011, Takeh2011} instead of an optimization problem, where we seek the \textit{best} TBF-CM that fits the data. This reformulation allows us to take an ensemble approach to make predictions using the CM, similar to how hurricane trajectories are forecast in weather modeling.\cite{Shanbhag2012, Krishnamurti2000}

\section{Conclusions}
\label{sec:conclusions}

The goal of this work is the discovery of parsimonious, physics-constrained CMs from OS measurements. Physical constraints like symmetry and frame-invariance enable us to embed these CMs into CFD software to predict complex flows in process equipment and accelerate design and scale-up of new products. This work builds on (i) tensor basis functions, which allow us to express unknown nonlinear terms in CMs using a finite number of basis functions, (ii) sparse regression techniques, which allow us to look for parsimonious CMs, and (iii) a computer program called FLASH, which allows us to efficiently evaluate the PSS solution of arbitrary nonlinear differential CMs in the PI scenario.

We generate synthetic data using the Giesekus and PTT models and considered two different scenarios, CI and PI. We assume that $N_1$, $N_2$, and $\sigma_{12}$ are measured in the CI scenario, while only $\sigma_{12}$ is measured in the PI scenario. The CI scenario leads to an SLR problem which can be efficiently solved using LASSO regression. For the more typical PI scenario, we propose a two-stage greedy SNLR algorithm. In the first stage, we identify the $k$ most promising (unknown) coefficients before optimizing them during the second stage.

In the case of the Giesekus model, both the SLR and the SNLR find the `true' model, due to the simple representation of the CM in terms of the TBFs. In the case of the PTT model, the TBF-CMs discovered in both the CI and PI scenarios fit the experimental data well. In OS, the predictive ability of the discovered TBF-CMs under operating conditions that interpolate the training data is nearly perfect. The CMs also extrapolate satisfactorily when extended to test conditions outside the convex hull of the training data. The out-of-sample performance of these TBF-CMs in the startup of steady and uniaxial extension tests, while not perfect, is still quite satisfactory. This work has important implications for the design of experiments, the necessity of exploring non-shear flows, and potential algorithmic improvements for SNLR.

\appendix

\section{Cayley-Hamilton Theorem and Tensor Invariants}
\label{app:cht}

The characteristic polynomial of any $3 \times 3$ square matrix $\bm{A}$ is,
\begin{equation}
p_A(\lambda) = \det (\bm{A}-\lambda \bm{I}),
\end{equation}
where $\det (\bm{A})$ denotes the determinant of matrix $\bm{A}$. The roots of $p_A(\lambda) = 0$ yield the eigenvalues ($\bm{\lambda} = [\lambda_1, \lambda_2, \lambda_3])$ of $\bm{A}$. A function $f(\bm{A})$ is an invariant of $\bm{A}$ if its value is independent of the rotation of $\bm{A}$. Mathematically,
\[f(\bm{A}) = f(\bm{QAQ}^T)\, \quad \quad \forall \text{ orthogonal matrices } \bm{Q}. \]
The three principal invariants of $\bm{A}$ are given by,
\begin{align}
  I_1 &= \mathrm{tr}(\bm{A}) = A_{11}+A_{22}+A_{33} = \lambda_1+\lambda_2+\lambda_3 \\   
  I_2 &= \frac{1}{2} \left( (\mathrm{tr} \left( \bm{A} \right) )^2 - \mathrm{tr}(\bm{A}^2) \right)   = \lambda_1 \lambda_2 + \lambda_1 \lambda_3 + \lambda_2 \lambda_3 \\ 
I_3 &= \det (\bm{A}) = \lambda_1 \lambda_2 \lambda_3.
\end{align}
The coefficients of the characteristic polynomial turn out to be the principal invariants
\begin{equation}
p_A(\lambda) = \lambda^3 - I_1 \lambda^2 + I_2 \lambda - I_3.
\end{equation}

The Cayley-Hamilton theorem states that the characteristic polynomial of \textit{any} square matrix evaluated at $\bm{A}$ equals zero,\cite{Cayley1858, Frobenius1877}
\begin{equation}
p_A(\bm{A}) = \bm{A}^3 - I_1 \bm{A}^2 + I_2 \bm{A} - I_3 \bm{I} = \bm{0}.
\label{eqn:cht1}
\end{equation}

It enables us to write $\bm{A}^n$ with $n \geq 3$ in terms of $\bm{A}^2$, $\bm{A}$, and $\bm{I}$ by recursively invoking Equation \ref{eqn:cht1}. For example,
\begin{align*}
\bm{A}^4 & = \bm{A}^3 \bm{A} = (I_1 \bm{A}^2 - I_2 \bm{A} + I_3 \bm{I}) \bm{A} \\
& = I_1 \bm{A}^3 - I_2 \bm{A}^2 + I_3 \bm{A} \\
& = I_1 (I_1 \bm{A}^2 - I_2 \bm{A} + I_3 \bm{I}) - I_2 \bm{A}^2 + I_3 \bm{A} \\
& = (I_1^2 - I_2) \bm{A}^2 + (I_3 - I_1 I_2) \bm{A} + (I_1 I_3) \bm{I}.
\end{align*}
This has an important corollary: the Taylor series expansion  of any analytic function $f(\bm{A})$ can be truncated after the quadratic term,
\begin{equation}
f(\bm{A}) = \sum_{k=0}^{\infty} c_k \bm{A}^k = g_2(I_1, I_2, I_3) \bm{A}^2 + g_1(I_1, I_2, I_3) \bm{A} + g_0(I_1, I_2, I_3) \bm{I}
\label{eqn:ch_func}
\end{equation}
where $\{g_i\}_{i=1}{3}$ are polynomial functions of the invariants, and hence the eigenvalues of $\bm{A}$.

\section{TBFs and Invariants in Shear}
\label{app:os_tbf}

In oscillatory or steady shear experiments, $\dot{\bm{\gamma}} = \dot{\gamma} \bm{e}_{12} + \dot{\gamma} \bm{e}_{21}$ and $\bm{\sigma} = \sigma_{11} \bm{e}_{11} + \sigma_{12} \bm{e}_{12} + \sigma_{12} \bm{e}_{21} + \sigma_{22} \bm{e}_{22}$. The TBFs listed in Table \ref{tab:tbf_and_invariant} can be further simplified by substituting these expressions into the general equations. They yield:
\begin{align*}
\bm{T}_1 & = 1 \bm{e}_{11} + 1 \bm{e}_{22} + 1 \bm{e}_{33}\\
\bm{T}_2 & = \sigma_{11}\bm{e}_{11} + \sigma_{12}\bm{e}_{12} + \sigma_{12}\bm{e}_{21} + \sigma_{22}\bm{e}_{22}\\
\bm{T}_3 & =  \dot{\gamma}\bm{e}_{12} + \dot{\gamma}\bm{e}_{21}\\
\bm{T}_4 & = \sigma_{11}^{2} + \sigma_{12}^{2}\bm{e}_{11} + \sigma_{12} \left(\sigma_{11} + \sigma_{22}\right)\bm{e}_{12} + \sigma_{12} \left(\sigma_{11} + \sigma_{22}\right)\bm{e}_{21} + \sigma_{12}^{2} + \sigma_{22}^{2}\bm{e}_{22}
\\
\bm{T}_5 & =  \dot{\gamma}^{2}\bm{e}_{11} + \dot{\gamma}^{2}\bm{e}_{22}
\\
\bm{T}_6 & = 2 \dot{\gamma} \sigma_{12}\bm{e}_{11} + \dot{\gamma} \left(\sigma_{11} + \sigma_{22}\right)\bm{e}_{12} + \dot{\gamma} \left(\sigma_{11} + \sigma_{22}\right)\bm{e}_{21} + 2 \dot{\gamma} \sigma_{12}\bm{e}_{22}\\
\bm{T}_7 & = 2 \dot{\gamma} \sigma_{12} \left(\sigma_{11} + \sigma_{22}\right)\bm{e}_{11} + \dot{\gamma} \left(\sigma_{11}^{2} + 2 \sigma_{12}^{2} + \sigma_{22}^{2}\right)\bm{e}_{12} + \dot{\gamma} \left(\sigma_{11}^{2} + 2 \sigma_{12}^{2} + \sigma_{22}^{2}\right)\bm{e}_{21} + \\
& \qquad \qquad 2 \dot{\gamma} \sigma_{12} \left(\sigma_{11} + \sigma_{22}\right)\bm{e}_{22}\\
\bm{T}_8 & = 2 \dot{\gamma}^{2} \sigma_{11}\bm{e}_{11} + 2 \dot{\gamma}^{2} \sigma_{12}\bm{e}_{12} + 2 \dot{\gamma}^{2} \sigma_{12}\bm{e}_{21} + 2 \dot{\gamma}^{2} \sigma_{22}\bm{e}_{22}
\\
\bm{T}_9 & = 2 \dot{\gamma}^{2} \left(\sigma_{11}^{2} + \sigma_{12}^{2}\right)\bm{e}_{11} + 2 \dot{\gamma}^{2} \sigma_{12} \left(\sigma_{11} + \sigma_{22}\right)\bm{e}_{12} + 2 \dot{\gamma}^{2} \sigma_{12} \left(\sigma_{11} + \sigma_{22}\right)\bm{e}_{21} + \\
& \qquad \qquad  2 \dot{\gamma}^{2} \left(\sigma_{12}^{2} + \sigma_{22}^{2}\right)\bm{e}_{22}.
\end{align*}
The corresponding invariants can be written as,
\begin{align*}
I_1 & = \sigma_{11} + \sigma_{22}\\
I_2 & = \sigma_{11}^{2} + 2 \sigma_{12}^{2} + \sigma_{22}^{2}
\\
I_3 & = 2 \dot{\gamma}^{2}\\
I_4 & = \sigma_{11}^{3} + 3 \sigma_{11} \sigma_{12}^{2} + 3 \sigma_{12}^{2} \sigma_{22} + \sigma_{22}^{3}\\
I_5 & = 0\\
I_6 & = 2 \dot{\gamma} \sigma_{12}\\
I_7 & = I_1 I_6\\
I_8 & = I_1 I_3/2\\
I_9 & = I_2  I_3/2.
\end{align*}
Thus, only five of the nine invariants ($I_1$, $I_2$, $I_3$, $I_4$, $I_6$) are nontrivial. The remaining invariants are zero, or products of other invariants. Thus, the task of learning $g(\bm{L})$ simplifies to the task of learning $\bm{g}(\bm{l})$, with $\bm{l} = [I_1, I_2, I_3, I_4, I_6]$.

\section*{Supplementary Material}
See supplementary material online for (i) nomenclature and abbreviations, (ii) Giesekus and PTT models in oscillatory shear, (iii) harmonic balance and FLASH, (iv) Giesekus model in the PI Scenario, (v) Extrapolation of TBF-CM in OS flow, (vi) TBFs and invariants in uniaxial extension.

\section*{Acknowledgments}

This work is based in part on work supported by the National Science Foundation under grant no. NSF DMR-1727870 (SS). The authors thank Kyle R. Lennon and Alexander Peterson for helpful discussions.

\section*{Data Availability Statement}

Data supporting the findings of this study are available from the corresponding author upon reasonable request.

\bibliography{extract}

\end{document}


\begin{center}
\textbf{\huge{Supplementary Material}}\\
\vspace{0.5cm}
\textbf{\Large Sparse Regression for Discovery of Constitutive Models from Oscillatory Shear Measurements}\\
\vspace{0.5cm}

\medskip

Sachin Shanbhag$^{*}$ and Gordon Erlebacher\\ 

Department of Scientific Computing, Florida State University, Tallahassee, Florida, USA

\end{center}

\vspace{0.5cm}

\tableofcontents

\newpage


\section{Nomenclature and Abbreviations}
\label{sec:nomenclature}

\begin{longtable}{p{.20\textwidth}  p{.85\textwidth} } 
        \hline
         \textbf{Abbreviation} & \textbf{Description}
         \\
         \hline
         CFD & Computational Fluid Dynamics \\
         CI & Complete Information \\
         CM & Constitutive Model \\

         FLASH & Fast Large Amplitude Simulation using Harmonic balance \\
         FFT & Fast Fourier Transform \\

         HB & Harmonic Balance \\
         IVP & Initial Value Problem \\

         LAOS & Large Amplitude Oscillatory Shear \\
         LASSO & Least Absolute Shrinkage and Selection Operator\\
         LS & Least Squares\\
         ML & Machine Learning\\
         NN & Neural Networks\\

         OS & Oscillatory Shear \\

         PF & Polynomial Feature (element of $\bm{B}_d$)\\
         PI & Partial Information \\
         PINNs & Physics Informed Neural Networks\\
         PSS & Periodic Steady State\\
         PTT & Phan-Thien Tanner \\
         RUDE & Rheological Universal Differential Equations\\

         SCF & Scalar Coefficient Function ($g_i$)\\
         SLR & Sparse Linear Regression \\
         SNLR & Sparse Nonlinear Regression \\
         SSR & Sum of Squared Residuals\\

		 TBF & Tensor Basis Function\\
         UCM & Upper Convected Maxwell \\
         \hline
    \caption{List of abbreviations used in the paper}
    \label{tab:abbr}
\end{longtable}

\subsection{Conventions}

Throughout the paper, bold symbols indicate vectors or tensors: bold lower-case symbols ($\bm{x}$) denote vectors or lists, while bold upper-case symbols ($\bm{X}$) denote tensors or matrices. A `dot' represents a rate of change or partial derivative, $\dot{x} = \partial x/\partial t$. A `hat' represents a Fourier transform: $\hat{\bm{x}}$ is a vector of Fourier coefficients of a periodic signal $x(t)$.

\begin{longtable}{p{.10\textwidth}  p{.85\textwidth} } 
        \hline 
        \textbf{Symbol} & \textbf{Description} \\
        \hline
        $\bm{\alpha}$ & vector of coefficients of SCFs in TBF-CM\\
        $\bm{\alpha}^{*}$ & optimal $\bm{\alpha}$ from fitting TBF-CM to experiments\\
        $\tilde{\bm{\alpha}}_j$ & $\bm{\alpha}$ where only the $j$th element is nonzero (one hot encoding)\\
	$\alpha_G$ & nonlinear parameter in the Giesekus model\\

	$\bm{B}_d$ & set of polynomial features for degree of approximation $d$\\
	$B_j$ & $j$th polynomial feature in the set $\bm{B}_d$\\

        $\gamma_0$ & strain amplitude \\
		$\gamma$ & shear strain \\
        $\dot{\bm{\gamma}}$ & deformation gradient tensor\\

        $d$ & degree of polynomial approximation of TBF-CM\\
	$\varepsilon_\text{PTT}$ & nonlinear parameter in the exponential PTT model\\
	$\dot{\varepsilon}$ & steady extensional strain rate in uniaxial extension\\
	$\bm{F}$ & nonlinear frame-invariant tensor function in the generalized UCM model \\
	$\Gp$ & storage modulus\\
	$\Gpp$ & loss modulus\\
        $G$ &  shear modulus\\
	$g_i$ & $i$th scalar coefficient function\\
	$\bm{g}$ & set of 9 SCFs ($=\{g_1, \cdots, g_9\})$\\

	$H$ & parameter that determines the highest harmonic ($2H + 1$) resolved in FLASH\\

	$I_i$ & $i$th invariant of $\sigma$  and $\dot{\bm{\gamma}}$ (Table \ref{m-tab:tbf_and_invariant})\\

	$\bm{L}$ & set of all 9 invariants of $\bm{\sigma}$  and $\dot{\bm{\gamma}}$ (=$I_1, \cdots, I_9$)\\
	$\bm{l}$ & $n_l$ = 5 independent invariants in OS $= [I_1, I_2, I_3, I_4, I_6] \subset \bm{L}$\\

        $N_1$ & first normal stress difference ($=\sigma_{11} - \sigma_{22}$)\\
        $N_2$ & second normal stress difference ($=\sigma_{22} - \sigma_{33}$)\\       

        $n_\alpha$ & number of elements of $\bm{\alpha}$\\
        $n_b$ & number of polynomial features  \\
        $n_\text{ds}$ & number of training datasets\\

        $n_g$ & number of scalar coefficient functions ($=9$) \\
        $n_l$ & number of independent invariants in oscillatory shear ($=5$) \\
        $n_{F_v}$ & number of processed observations in CI scenario\\
      
        $n_\sigma$ & number of components of $\bm{\sigma}$ resolved in FLASH ($=3$) \\
        $n_t$ & number of temporal grid points over one cycle ($=2^{6}$) \\

	$P_k$ & list of indices of the $k$ most promising coefficients\\

        $\bm{\sigma}$ & extra stress tensor\\
        $\hat{\bm{\sigma}}_\text{exp}$ & Fourier transform of experimental $\bm{\sigma}$ data\\     

        $\sigma_{12}$ & shear stress \\
        $\sigma_{11}, \sigma_{22}, \sigma_{33}$ & normal stresses \\

        $\stackrel{\triangledown}{\bm{\sigma}}$ & upper-convected derivative of tensor $\bm{\sigma}$ \\

	$\bm{T}_i$ & $i$th tensor basis function of $\bm{\sigma}$  and $\dot{\bm{\gamma}}$ (Table \ref{m-tab:tbf_and_invariant})\\

	$\mathbb{T}$ & the set of TBFs, ( = $\{\bm{T}_i\}_{i=1}^{9}$)\\
	
        $\tau$ & relaxation time\\

	$\chi^2$ & time-domain loss function for least squares regression\\
	$\chi^2_{\text{LASSO}}$ & time-domain loss function for LASSO regression \\
		$\chi^2_F$ & frequency-domain loss function for the PI scenario\\
    $\omega$ & angular frequency \\	

	\hline
    \caption{List of symbols used in the paper}
    \label{tab:nomenclature}
\end{longtable}


\clearpage

\section{Giesekus and PTT models in Oscillatory Shear}
\label{sec:ode_models}

\subsection{Giesekus model}
The Giesekus model is given by,
\begin{align}
\dfrac{d \bm{\sigma}}{d t} & - \bm{\nabla v}^{T} \cdot \bm{\sigma} - \bm{\sigma} \cdot \bm{\nabla v} + \dfrac{1}{\tau} \bm{\sigma} + {\color{blue} \bm{F}_\text{Giesekus}(\bm{\sigma},  \dot{\bm{\gamma}})} = G  \dot{\bm{\gamma}}\\
= \dfrac{d \bm{\sigma}}{d t} & - \bm{\nabla v}^{T} \cdot \bm{\sigma} - \bm{\sigma} \cdot \bm{\nabla v} + \dfrac{1}{\tau} \bm{\sigma}  + {\color {blue} \dfrac{\alpha_G}{G \tau} \bm{\sigma} \cdot \bm{\sigma}} = G  \dot{\bm{\gamma}},
\label{eqn:giesekus_tensor}
\end{align}

In OS, Eq. \eqref{eqn:giesekus_tensor} simplifies to the set of nonlinear ordinary differential equations ,
\begin{align}
\dot{\sigma}_{11} & + \frac{\sigma_{11}}{\tau}+\frac{\alpha_G}{G \tau}\left(\sigma_{11}^{2}+\sigma_{12}^{2}\right)-2\dot{\gamma}\sigma_{12}=0 \notag\\
\dot{\sigma}_{22} & +\frac{\sigma_{22}}{\tau}+\frac{\alpha_G}{G \tau}\left(\sigma_{22}^{2}+\sigma_{12}^{2}\right)=0 \notag\\
\dot{\sigma}_{33} & +\frac{\sigma_{33}}{\tau}+\frac{\alpha_G}{G \tau}\sigma_{33}^{2}=0 \notag \\
\dot{\sigma}_{12} & +\frac{\sigma_{12}}{\tau}+\frac{\alpha_G}{G \tau}\left(\sigma_{11}+\sigma_{22}\right)\sigma_{12}-\sigma_{22}\dot{\gamma}-G\dot{\gamma}=0.
\label{eq:giesekus_ode}
\end{align}

\subsection{PTT model}

The exponential PTT model is given by,
\begin{align}
\dfrac{d \bm{\sigma}}{d t} & - \bm{\nabla v}^{T} \cdot \bm{\sigma} - \bm{\sigma} \cdot \bm{\nabla v} + \dfrac{1}{\tau} \bm{\sigma} + {\color {blue} \bm{F}_\text{PTT}(\bm{\sigma},  \dot{\bm{\gamma}})} = G  \dot{\bm{\gamma}}\\
= \dfrac{d \bm{\sigma}}{d t} & - \bm{\nabla v}^{T} \cdot \bm{\sigma} - \bm{\sigma} \cdot \bm{\nabla v} + \dfrac{f_\text{PTT}(\bm{\sigma})}{\tau} \bm{\sigma} = G  \dot{\bm{\gamma}},
\label{eqn:ptt_tensor}
\end{align}
where $\varepsilon_\text{PTT} \in [0, 1]$ controls the nonlinear term via,
$$f_\text{PTT}(\bm{\sigma} ) = \exp\left(\dfrac{\varepsilon_\text{PTT}}{G} \text{tr} (\bm{\sigma}) \right).$$

In OS, the set of nonlinear ordinary differential equations is given by,
\begin{align}
\dot{\sigma}_{11} & + \frac{f_\text{PTT}(\bm{\sigma} )}{\tau}\sigma_{11}  - 2\dot{\gamma} \sigma_{12} = 0 \nonumber\\
\dot{\sigma}_{22} & + \frac{f_\text{PTT}(\bm{\sigma} )}{\tau}\sigma_{22} = 0 \nonumber\\
\dot{\sigma}_{33} & + \frac{f_\text{PTT}(\bm{\sigma} )}{\tau}\sigma_{33} = 0. \nonumber\\
\dot{\sigma}_{12} & + \frac{f_\text{PTT}(\bm{\sigma} )}{\tau}\sigma_{12} - \dot{\gamma} \sigma_{22} - G \dot{\gamma} = 0.
\label{eqn:ptt_ode}
\end{align}

\section{Harmonic Balance and FLASH}
\label{sec:hb_flash}

Harmonic balance (HB) is a general numerical method for solving systems of nonlinear ODEs with oscillatory forcing.\cite{Krack2019} The simplest way to introduce the idea of HB to solve CMs under OS is to consider a first-order ordinary differential equation (ODE) in a single dependent variable $q(t)$ subjected to sinusoidal external forcing $f_{\text{ex}}(t) = f_0 \sin \omega t$,
\begin{equation}
\dot{q}(t) + f_{\nl}(q,t) -  f_{\ex}(t) = 0.
\label{eqn:general_form}
\end{equation}
Here $f_{\nl}(q,t)$ is a (potentially) nonlinear function of $q$ (e.g., $f_{\nl}(q,t) = q^2$), and $q(t) = q(t + T)$ is a periodic function with period $T = 2\pi/\omega$. A standard approach for solving the Eq. \ref{eqn:general_form} is to use a time-stepping integration method like Runge-Kutta with some initial condition (say, $q(0) = 0$), wait for transients to decay and a PSS solution to emerge. Instead, HB starts by specifying an \textit{ansatz} for the periodic steady state (PSS) solution $q(t)$ in terms of a truncated Fourier series expressed as either trigonometric $\left(\sin{k\omega t} \text{ and } \cos{k\omega t}\right)$ or complex $(e^{ik\omega t})$ basis functions. The Fourier series representation of $q(t)$ up to $H$ harmonics is, 
\begin{equation}
q(t)\approx q_{H}(t)=\sum_{k=-H}^{H} \hat{q}(k) \,e^{ik\omega t},
\label{eq:FourierSeries}
\end{equation}
where $\hat{q}(k)$ is the complex Fourier coefficient associated with the $k$th harmonic, defined via
\begin{equation}
\hat{q}(k) = \dfrac{1}{T} \int_{0}^{T} q(t)\,e^{-ik\omega t} dt.\label{eq:FourierTransform}
\end{equation}
The number of Fourier coefficients in Eq. \eqref{eq:FourierSeries} is $2H + 1$.

\medskip

In HB, instead of solving Eq. \ref{eqn:general_form}, we consider its projection in the ansatz, i.e., $\dot{q}_H(t) + f_{\nl}(q_H,t) -  f_{\ex}(t) = 0$. It can be shown that the HB equations correspond to the weak form,
\begin{equation}
\hat{\dot{q}}_H(k) + \hat{f}_{\nl}(k, \hat{q}_H) -  \hat{f}_{\ex}(k) = 0, \quad k = -H, \cdots, H.
\label{eqn:hb_eqns}
\end{equation} 
This is a system of $2H+1$ nonlinear equations in the $2H+1$ unknown coefficients $\hat{q}_H(k)$.

\medskip

Since external forcing is prescribed, $\hat{f}_{\ex}(k)$ can be determined analytically. Using Eq. \eqref{eq:FourierSeries}, it can be shown that the Fourier transform of the derivative
$\hat{\dot{q}}_H(k) = ik\omega \, \hat{q}(k)$. Except in special situations, the Fourier transform $\hat{f}_{\nl}$ cannot be determined analytically. The alternating time frequency (AFT) scheme approaches this problem numerically. It is a versatile and efficient method that evaluates $\hat{f}_{\nl}$ from $\hat{q}$ using a combination of inverse FFT and FFT calculations. It involves three steps: 
\begin{enumerate}[(i)]
\item $\hat{q}(k) \rightarrow q_H(t)$: use inverse FFT to obtain $q_H(t)$ on a uniform grid with $n_t$ points, where $n_t > 2(2H + 1)$ to avoid aliasing error;\cite{heath2018scientific} 
\item $q_H(t) \rightarrow f_{\nl}(q_H, t)$: substitute $q_H(t)$ into the specific expression for the nonlinear term $f_{\nl}(q,t)$ in terms of $q$ (e.g., $f_{\nl}(t_i) = q(t_i)^2$ for $i \in [1, n_t]$);
\item  $f_{\nl} (t) \rightarrow \hat{f}_{\nl} $: use FFT to obtain $\hat{f}_{\nl}(k)$ for $-H \leq k \leq H$.
\end{enumerate}
The $\hat{f}_{\nl}$ obtained using AFT is used in each iteration while solving Eq. \eqref{eqn:hb_eqns} using a nonlinear solver.  The computational cost of FFT and its inverse is $\mathcal{O}(n_t \log n_t)$.


\subsection{FLASH}

FLASH uses HB with AFT to solve differential CMs subjected to oscillatory shear. Due to symmetry, only even harmonics are present in the normal stresses $\sigma_{11}$ and $\sigma_{22}$, while only odd harmonics are present in the shear stress $\sigma_{12}$. In this ansatz,\cite{Mittal2023} 
\begin{align}
\sigma_{11}(t) & \approx \sum_{k=-H}^{H} \hat{\sigma}_{11}(2k) e^{i2k\omega t} \notag\\
\sigma_{22}(t) & \approx \sum_{k=-H}^{H} \hat{\sigma}_{22}(2k) e^{i2k\omega t} \notag\\
\sigma_{12}(t) & \approx \sum_{k=-(H+1)}^{H} \hat{\sigma}_{12}(2k+1) e^{i(2k+1)\omega t}.
\label{eqn:ansatz}
\end{align}
Each complex Fourier coefficient has a real and imaginary part. Since all stresses are real, the Fourier coefficients are complex conjugates, i.e., $\hat{a}(2k)=\bar{\hat{a}}(-2k)$, where the overbar is used to denote the complex conjugate. Thus, we do not track the Fourier coefficients corresponding to $k < 0$. Furthermore, the Fourier coefficient corresponding to $k=0$, that is, $\hat{a}(0)$, is strictly real. As a result of these simplifications, $\hat{\sigma}_{11}$ and $\hat{\sigma}_{22}$ can be represented using $2H + 1$ real coefficients, while $\hat{\sigma}_{12}$ can be represented using $2H + 2$ real coefficients. Thus, the \textbf{highest harmonic resolved in FLASH} is $2H + 1$, and the \textbf{total number of real Fourier coefficients} is $2(2H+1) + 2H+2 = 6H + 4$.

\medskip

An implementation of FLASH using Python was included as supplementary online material in the paper that first described this method.\cite{Mittal2024a}. 

\newpage

\section{Giesekus Model in the PI Scenario}
\label{sec:Giesekus_PI}

Similarly to the PTT model discussed in the manuscript, we applied the SNLR algorithm to synthetic data obtained from the Giesekus model with $\alpha_G = 0.3$ at the same $n_\text{ds} = 12$ choices of $(\omega, \gamma_0)$. We set $d = 2$, which implies $n_{\alpha} = 189$. 

\medskip

\begin{figure}[h!]
\begin{center}
\includegraphics[scale=0.6]{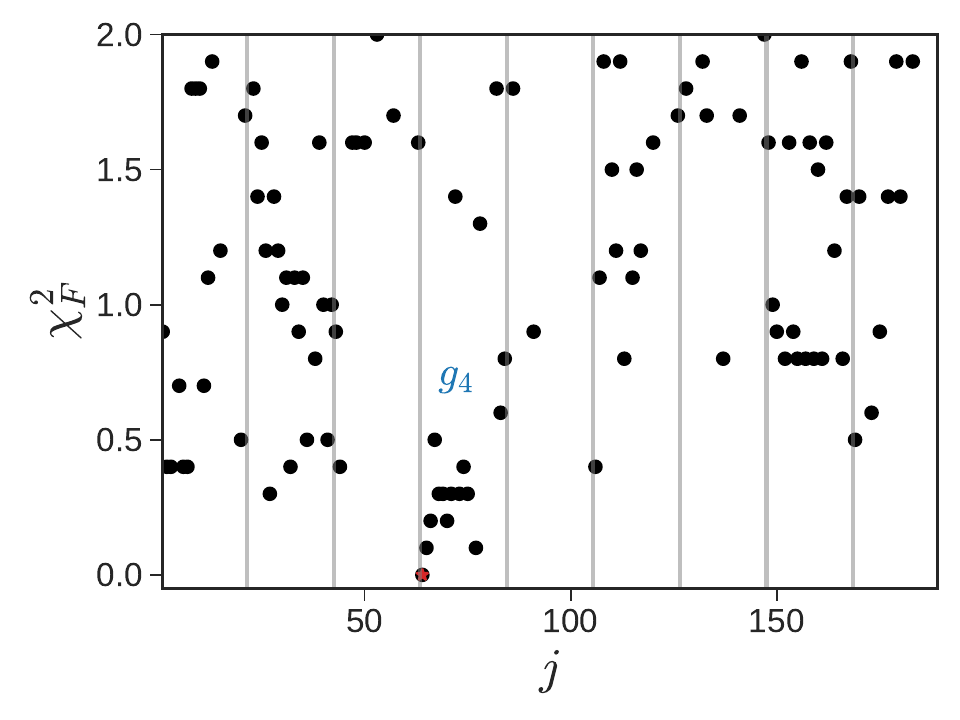}
\end{center}
\caption{\textbf{Giesekus Model in the PI Scenario}: Loss functions $\chi^2_F(\tilde{\bm{\alpha}}_j)$, where $j \in [1, n_\alpha = 189]$, obtained after the first stage of the SNLR algorithm. The red star corresponds to the smallest value of $\chi^2_F(\tilde{\bm{\alpha}}_j)$, and highlights the most promising coefficient ($\alpha_{4, 1}$). \label{fig:giesekus_optim1c}}
\end{figure}

Results of the first stage of the algorithm, where we scan $\tilde{\bm{\alpha}}_j$ for $j \in [1, n_{\alpha} = 189]$ by performing $n_\alpha$ independent 1D nonlinear minimizations of the loss function (Equation \eqref{m-eqn:loss_chiF}) to obtain the optimal values $\tilde{\bm{\alpha}}_j^{*}$ and the corresponding loss functions $\chi^{2}_{F}(\tilde{\bm{\alpha}}_j^{*})$ are shown in Figure \ref{fig:giesekus_optim1c}. 

The first stage not only correctly identifies the importance of $g_4$ for the Giesekus model, but picks out the ($k = 1$) most promising coefficient, $\alpha_{4, 1} \equiv \alpha_{64}$, which is indicated by a red star. The average cost of each of these minimizations was 22.4 $\pm$ 14.2 s.

\medskip

Performing a second stage is somewhat redundant, but nevertheless leads us to the true model,
$$\bm{F}_{k1}^\text{SNLR} = \bm{F}_\text{Giesekus} = 0.30\, \bm{T}_4.$$
Since the proposed SNLR algorithm discovers the true CM, we do not report its performance in different settings like we do with the PTT model in the manuscript.

%

\newpage

\section{Extrapolation of TBF-CM in OS Flow}
\label{sec:extrap_pi_os}

From Figure \ref{m-fig:predictTBF_OS}, we observe that the TBF-CMs inferred from the PTT data in the PI scenario extrapolate reasonably well. From our numerical experiments, the inferred TBF-CMs extrapolate reliably well beyond the training range as shown in Figure \ref{fig:os_extrap} below, where the smallest frequency ($\omega \tau = 0.01$) is a factor of ten smaller than the smallest frequency in the training data ($\omega \tau = 0.1$). Similarly, predictions of $\sigma_{12}$ at frequencies five times larger than those of the training set ($\omega \tau = 10$) also appear reasonable.

\begin{figure}[h!]
\begin{center}
\includegraphics[scale=0.6]{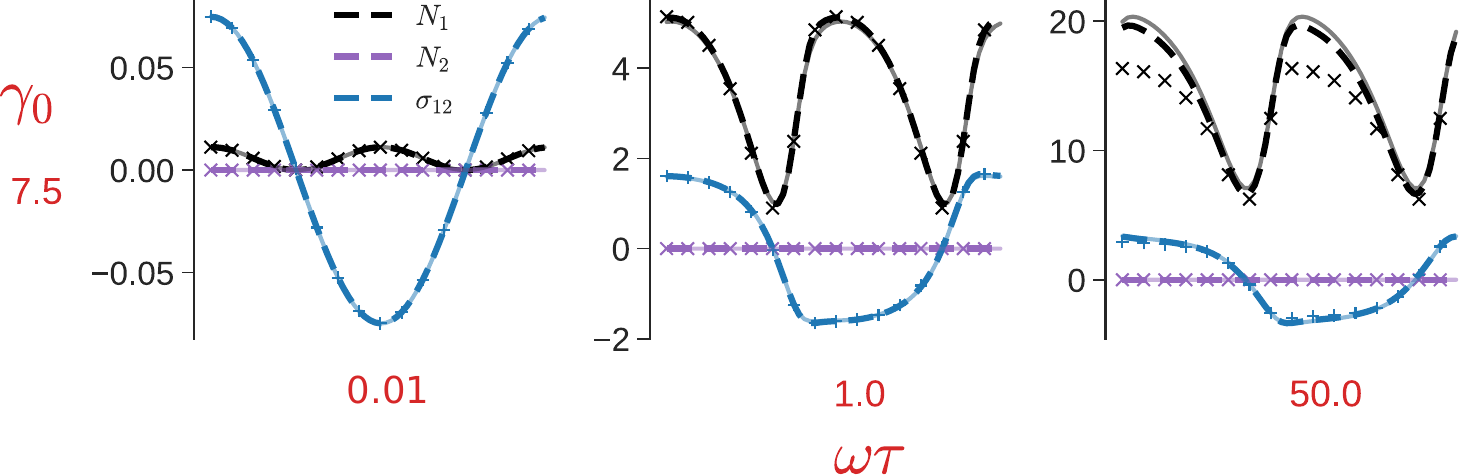}
\end{center}
\caption{\textbf{PI OS Flow Extrapolation}. Extrapolations of the two different TBF-CMs (Eq. \eqref{m-eqn:top3tbf}) are compared with the response of the PTT model with $\epsilon_\text{PTT} = 0.3$ using $H = 8$ and $n_t = 2^6$. The horizontal axis in each subfigure represents $\omega t \in [0, 2\pi]$. Symbols and lines have the same meaning as Figure \ref{m-fig:fitTBF_OS} of the manuscript. \label{fig:os_extrap}}
\end{figure}

For simulations with FLASH at $\omega \tau \gtrsim 50$ and/or $\gamma_0 \gtrsim 7.5$, it is recommended to use values of $H > 8$ to capture the effect of higher harmonics and to use either a larger number of intermediate steps or a better initial guess to aid convergence. 


\medskip

No special care is required for numerical integration. From our numerical experiments, while the accuracy of TBF-CMs for the PTT model suffers somewhat as we extrapolate far beyond the training data (evident in the $N_1$ predictions in Figure \ref{fig:os_extrap}), they do not fail catastrophically. Instead, they smoothly diverge from the true model.

\newpage

\section{TBFs and Invariants in Uniaxial Extension}
\label{sec:TBF_unaxial}

In uniaxial extension, $\bm{\nabla v} = \dot{\varepsilon}\, \bm{e}_{11} - 0.5 \dot{\varepsilon}\, \bm{e}_{22} - 0.5 \dot{\varepsilon}\, \bm{e}_{33}$, where $\dot{\varepsilon}$ is the steady extensional strain rate. This implies that the shear rate tensor $\dot{\bm{\gamma}} = 2 \bm{\nabla v} = 2 \dot{\varepsilon}\, \bm{e}_{11} - \dot{\varepsilon}\, \bm{e}_{22} - \dot{\varepsilon}\, \bm{e}_{33}$, and the stress tensor $\bm{\sigma} = \sigma_{11}  \bm{e}_{11} + \sigma_{22}  \bm{e}_{22} + \sigma_{33}  \bm{e}_{33}$ has three nonzero components along the diagonal. The TBFs listed in Table \ref{m-tab:tbf_and_invariant} of the manuscript can be further simplified by substituting these expressions into the general equations.
\begin{align*}
\bm{T}_1 & = 1 \bm{e}_{11} + 1 \bm{e}_{22} + 1 \bm{e}_{33}\\
\bm{T}_2 & = \sigma_{11}\bm{e}_{11} + \sigma_{22}\bm{e}_{22} + \sigma_{33} \bm{e}_{33}\\
\bm{T}_3 & =  2\dot{\varepsilon} \bm{e}_{11} - \dot{\varepsilon} \bm{e}_{22} - \dot{\varepsilon} \bm{e}_{33}\\
\bm{T}_4 & = \sigma_{11}^2 \bm{e}_{11} + \sigma_{22}^2 \bm{e}_{22} + \sigma_{33}^2 \bm{e}_{33}\\
\bm{T}_5 & =  4\dot{\varepsilon}^2 \bm{e}_{11} + \dot{\varepsilon}^2 \bm{e}_{22} + \dot{\varepsilon}^2 \bm{e}_{33}\\
\bm{T}_6 & =  4\dot{\varepsilon} \sigma_{11} \bm{e}_{11} - 2 \dot{\varepsilon} \sigma_{22} \bm{e}_{22} - 2 \dot{\varepsilon} \sigma_{33} \bm{e}_{33}\\
\bm{T}_7 & = 2 \dot{\varepsilon} \sigma_{11}^2 \bm{e}_{11} -  \dot{\varepsilon} \sigma_{22}^2 \bm{e}_{22} - \dot{\varepsilon} \sigma_{33}^2 \bm{e}_{33}\\
\bm{T}_8 & = 2 \dot{\varepsilon}^{2} \sigma_{11} \bm{e}_{11} + 0.5 \, \dot{\varepsilon}^2 \sigma_{22} \bm{e}_{22} + 0.5\, \dot{\varepsilon}^{2} \sigma_{33}^2 \bm{e}_{33}\\
\bm{T}_9 & = 2 \dot{\varepsilon}^2 \sigma_{11}^2 \bm{e}_{11} + 0.5 \, \dot{\varepsilon}^2 \sigma_{22}^2 \bm{e}_{22} + 0.5\, \dot{\varepsilon}^2 \sigma_{33}^2 \bm{e}_{33}.
\end{align*}
The corresponding invariants can be written as,
\begin{align*}
I_1 & = \sigma_{11} + \sigma_{22} + \sigma_{33}\\
I_2 & = \sigma_{11}^{2} + \sigma_{22}^{2} + \sigma_{33}^{2}
\\
I_3 & = 6 \dot{\varepsilon}^{2}\\
I_4 & = \sigma_{11}^{3} + \sigma_{22}^{3} + \sigma_{33}^{3}\\
I_5 & = 6 \dot{\varepsilon}^{3}\\
I_6 & = 2 \dot{\varepsilon} \sigma_{11} - \dot{\varepsilon} \sigma_{22} - \dot{\varepsilon} \sigma_{33} \\
I_7 & = \dot{\varepsilon} \left(2 \sigma_{11}^{2} - \sigma_{22}^{2} - \sigma_{33}^{2} \right)\\
I_8 & = \dot{\varepsilon} \left(4 \sigma_{11} + \sigma_{22} + \sigma_{33} \right)\\
I_9 & = \dot{\varepsilon} \left(4 \sigma_{11}^2 + \sigma_{22}^2 + \sigma_{33}^2 \right).
\end{align*}

\newpage

\printbibliography